\documentclass[letterpaper,twocolumn,preprintnumbers,amsmath,amssymb,floatfix,nofootinbib]{revtex4}

\usepackage{graphicx}
\usepackage{bm}
\usepackage{url,multirow,xcolor}

\begin{document}

\title{Examinations of the consistency of the quasiparticle random-phase approximation approach to double-$\bm{\beta}$ decay of $^{\bm{48}}$Ca }

\author{J.\ Terasaki}
\affiliation{Institute of Experimental and Applied Physics, Czech Technical University in Prague, Horsk\'{a} 3a/22, 128 00 Prague 2, Czech Republic}

\begin{abstract} 
The nuclear matrix elements (NMEs) of the neutrinoless and two-neutrino double-$\beta$ decays of $^{48}$Ca are calculated by the quasiparticle random-phase approximation (QRPA) with emphasis on the consistency examinations of this calculation method. The main new examination points are the consistency of two ways to treat the intermediate-state energies in the two-neutrino double-$\beta$ NME and comparison with the experimental charge-exchange strength functions obtained from $^{48}$Ca$(p,n)$ and $^{48}$Ti$(n,p)$ reactions. 
No decisive problem preventing the QRPA approach is found. The obtained neutrinoless double-$\beta$ NME adjusted by the ratio of the effective and bare axial-vector current couplings is lowest in those calculated by different groups and close to one of the QRPA values obtained by another group. 
\end{abstract}

\pacs{ }
%
\maketitle
\section{\label{sec:introduction}Introduction}
If the neutrinoless double-$\beta$ ($0\nu\beta\beta$) decay is observed, one can conclude that the neutrino is a Majorana particle. 
In this case, the effective neutrino mass can be determined by the half-life of the $0\nu\beta\beta$ decay, expected to be measured by the experiments, the phase-space factor, and the nuclear matrix element (NME). Recently the study of the $0\nu\beta\beta$ decay has obtained a stronger motivation than before by the discovery of the neutrino oscillation \cite{Fuk98,Ahm02,Egu03,Ali05} proving the existence of the finite neutrino mass. 
The phase-space factor and NME are the quantities that the theory should supply, and the latter is more difficult than the former because the accurate nuclear many-body wave functions are necessary. As is well-known, the calculated values of the NMEs are distributed in the range of a factor of 2$-$3 \cite{Eng17}, and this range is not reduced in spite of the effort of many theorists. For now there is no perfect calculation because all of the candidate nuclei for the $0\nu\beta\beta$ decay are heavy so that the exact nuclear wave functions cannot be obtained. In addition, effective strength of the spin-isospin transition operators is necessary for reproducing the related experimental data. 

One of the tasks that the theorists need to do is to examine the consistency of their calculations for clarifying the reliability. The purpose of this paper is to examine the consistency of the QRPA approach to the $\beta\beta$ decays of $^{48}$Ca in detail. There are two main check points not yet investigated. One is the treatment of the intermediate-state energy in the two-neutrino double-$\beta$ ($2\nu\beta\beta$) decay. The QRPA approach has two sets of the intermediate states defined by the QRPA calculations based on the initial and final states. I clarify the validity of using the two sets of intermediate-state energies  in the $2\nu\beta\beta$-NME calculation. Another new check point is the comparison of the Gamow-Teller (GT) strength function between the experimental data \cite{Yak09} and the calculation. This check point includes a question of whether theory can explain the quenching of the experimental GT strength. 

My motivation to investigate $^{48}$Ca is based on a fact that this mother nucleus is not always discussed in the papers of the systematic application of the QRPA to the $\beta\beta$ decays. I clarify in this paper whether $^{48}$Ca $\rightarrow$ $^{48}$Ti is particularly difficult for the QRPA approach. Several experimental projects searching for the $0\nu\beta\beta$ decay of $^{48}$Ca are in progress, or have been finished, see Refs.~\cite{Bru00} (TGV), 
\cite{Oga04} (ELEGANT VI), \cite{Bar11} (NEMO-3), \cite{Arn10} (SuperNEMO), \cite{Ume08} (CANDLES), \cite{Zde05} (CARVEL), and \cite{You91}. The advantage of $^{48}$Ca is the large Q value (4.7 MeV), and this nucleus is one of the major candidates for the $0\nu\beta\beta$ decay. Thus, it is worthy of  investigating theoretically in detail. 

My calculation method is explained in Sec.~\ref{sec:calculation_method} specifically for $^{48}$Ca $\rightarrow$ $^{48}$Ti including the technical aspects. The method to examine the use of two sets of the intermediate-state energies in the $2\nu\beta\beta$ NME is described in Sec.~\ref{sec:2vbb}. The results of the calculations are shown in Sec.~\ref{sec:calculation}, and the comparison with the results of other groups is made. The GT strength function is discussed in Sec.~\ref{sec:GT_strength_function}, and this study is summarized in Sec.~\ref{sec:summary}. 

\section{\label{sec:calculation_method}Calculation method}
\subsection{\label{sec:HFB}Hartree-Fock-Bogoliubov calculation}
$^{48}$Ca is not often discussed in the QRPA approach. 
This may be because the pairing gaps of the ground state of this nucleus are not as certain as those of other nuclei. I explain how the pairing gaps are determined in my calculation.  
The three-point formula \cite{Boh69} is used for obtaining the experimental pairing gaps from the experimental nuclear masses \cite{nndc}. 
The formula for the neutron pairing gap is 
\begin{align}
\bar{\Delta}_n = -\frac{1}{2}\left\{ \mathfrak{B}(N-1,Z) - 2\mathfrak{B}(N,Z) + \mathfrak{B}(N+1,Z) \right\}, 
\end{align}
where $\mathfrak{B}(N,Z)$ denotes the binding energy of the nucleus with the neutron number $N$ and proton number $Z$, and that for the protons, $\bar{\Delta}_p$, is obtained analogously. 
The presumption is that the odd-even mass staggering in the systematics occurs solely by the pairing correlations. Thus, this method is usually not used for the magic nuclei. The pairing gaps of $^{48}$Ti deduced from the masses are $\bar{\Delta}_p=2.343$ MeV and $\bar{\Delta}_n=1.742$ MeV. I reproduced approximately these pairing gaps by the Hartree-Fock-Bogoliubov (HFB) calculation  \cite{Bla05,Obe07,Ter03} using the Skyrme interaction SkM$^\ast$ \cite{Bar82} and the like-particle contact pairing interactions $\propto \delta(\bm{r}_1-\bm{r}_2)$ with the strengths adjusted for the protons and neutrons separately: $-258.4$ MeV$\,$fm$^3$ (protons) and $-224.5$ MeV$\,$fm$^3$ (neutrons) with the active range of the pairing interaction up to 30 MeV of the effective single-particle energy \cite{Bla05}. 
The pairing gaps obtained by my calculation are  
$\Delta_p=2.200$ MeV and $\Delta_n=1.671$ MeV (average pairing gaps). I also use this pairing interaction for $^{48}$Ca assuming that the pairing-interaction strength does not change significantly by the small change in the proton and neutron numbers. 
In order to check this assumption, I performed the HFB calculation for $^{44}$Ar and obtained 
$\Delta_p=2.080$ MeV and $\Delta_n=1.792$ MeV. The corresponding experimental values are 
$\bar{\Delta}_p=2.285$ MeV and $\bar{\Delta}_n=1.783$ MeV. This result justifies my procedure to treat the pairing interaction.

Using this pairing interaction, I obtained 
$\Delta_p=1.731$ MeV and $\Delta_n=0$ MeV for $^{48}$Ca. The neutrons have more difficulty in getting the pairing gap than the protons because there are shell gaps above and below the neutron Fermi surface at the first $f_{7/2}$ (1$f_{7/2}$) orbital. 
Let us see the low-lying spectra for seeking possible reflection of the proton pairing gap. The first experimental excited state of $^{48}$Ca is at 3.831 MeV \cite{nndc} with $J^\pi=2^+$. Since the proton one-particle-one-hole (1p-1h) excitation with the positive parity needs so called the  $2\hbar\omega$ jump, the main components of the excitation are those of the neutrons in the QRPA state. 
The second experimental excited state of $^{48}$Ca is at 4.283 MeV ($0^+$). Again the 1p-1h excitations with the  $2\hbar\omega$ jump are necessary for having the $0^+$ excited state in the QRPA. This condition applies for both the protons and neutrons, so that the QRPA cannot create that low-lying $0^+$ state. 
The third and fourth excited states are at 4.503 MeV ($4^+$) and 4.506 MeV ($3^-$), respectively, and the corresponding QRPA excitation energies are 3.956 MeV ($4^+$) and 5.550 MeV ($3^-$). No clear indication is obtained on the proton pairing gap from the corresponding energies. 
If enhancement of the two-proton transfer is seen experimentally, it would be the encouraging indication. However, there is no experiment of that reaction for $^{48}$Ca currently. Below I use the HFB solutions with the finite proton pairing gap but no neutron one. The uncertainty of the pairing gaps of $^{48}$Ca is minimized by the self-consistent calculation of the HFB approximation. 

\subsection{\label{sec:QRPA}QRPA calculation and technical parameters}
The calculation scheme of the QRPA is the same as that used in Refs.~\cite{Ter15,Ter16}. Here I note some technical parameters related to the accuracy of the calculation. 
The single-particle basis for representing the QRPA Hamiltonian matrix is constructed by the diagonalization of the one-body density matrix obtained from the HFB solutions (the canonical basis \cite{Rin80}) with the axial and parity symmetries (the symmetry axis is $z$). This basis is identical to the HF basis, if there is no pairing gap. 
The number of the single-particle states used in the calculations of this paper is around 1600$-$1700 including those with both the positive and negative $j_z$ (the $z$-component of the angular momentum) for each of the protons and neutrons. The maximum $j_z$ is 19/2. That dimension of the single-particle space approximately corresponds to $15\hbar\omega$ harmonic-oscillator shells. The wave functions are expressed with the B-spline mesh \cite{Bla05,Obe07,Ter03} in a cylinder box with the radius of 20 fm in the $xy$ plane and $0 \leq z \leq 20$ fm. The root-mean-square radius of $^{48}$Ca is 3.531 fm in the HFB solution. The number of mesh points is 42 for the region of 20 fm. The spherical symmetry of the spherical nuclei can be satisfied accurately with this geometrical preparation. It was confirmed by the HFB calculation that the ground states of $^{48}$Ca and $^{48}$Ti are spherical.  Many of the single-particle states are in the discretized-continuum region. 
The  density matrix and pairing tensor are calculated using the HFB wave functions in the active energy range mentioned in the previous section.  

The dimension of the two-quasiparticle basis for representing the QRPA excitation is truncated by the cutoff scheme  used in the previous calculations \cite{Ter15,Ter16}. The cutoff criteria in those calculations were determined so as to obtain the convergence of the final NMEs with respect to the dimension of the two-quasiparticle space and to satisfy  the geometrical symmetries of the Hamiltonian accurately in the calculation. The same criteria are used for the calculation of $^{48}$Ca and $^{48}$Ti of this paper. That dimension decreases as the $K$ quantum number (total $j_z$ of the nucleus) increases; it is approximately 24000 for $K^\pi=0^+$ and 13000 for $K^\pi=7^+$.  

\section{\label{sec:2vbb} $\bm{2\nu\beta\beta}$ nuclear matrix element }
The $2\nu\beta\beta$ NME, see Eq.~(25)$-$(28) in Ref.~\cite{Ter16}, can be written 
\begin{align}
M^{(2\nu)} =& \frac{M^{(2\nu)}_{GT}}{\mu_0} - \frac{g_V^2}{g_A^2} \frac{M^{(2\nu)}_F}{\mu_{0F}}, \label{eq:2vbb_NME}
\end{align}
with the $2\nu\beta\beta$ GT NME $M^{(2\nu)}_{GT}/\mu_0$, the $2\nu\beta\beta$ Fermi NME $M^{(2\nu)}_F/\mu_{0F}$, the vector-current coupling $g_V$, and the axial-vector current coupling $g_A$. The formulation for the axially-symmetric nuclei is applied to the spherical nuclei $^{48}$Ca and $^{48}$Ti in the calculation. The initial and final states of the $\beta\beta$ decay are $J^\pi=0^+$ states. Under these conditions, 
the $2\nu\beta\beta$ GT NME is given by 
\begin{align}
\frac{ M^{(2\nu)}_{GT} }{ \mu_0 } =& 
\ 3 \sum_{a^{K=0}_\textrm{exa}} \frac{1}{\mu_a} \langle F_\textrm{exa} | \tau^- \sigma_{K=0} | a^{K=0}_\textrm{exa}\rangle \nonumber \\
&\times\langle a^{K=0}_\textrm{exa} | \tau^- \sigma_{K=0} | I_\textrm{exa} \rangle  \label{eq:nme2vGT_exa} \\
=& 
\ 3 \langle F_\textrm{exa} | \tau^- \sigma_{K=0} \frac{m_e c^2}{ H - \bar{M} } \tau^- \sigma_{K=0} | I_\textrm{exa} \rangle,  \label{eq:nme2vGT_exa2}
\end{align}
\begin{align}
\mu_a =& \frac{1}{m_e c^2} ( E_{a, \textrm{exa}}^{K=0} - \bar{M} ), \label{eq:mu_a} \\
\bar{M} =& \frac{1}{2}( M_I + M_F ). \label{eq:bar_M}
\end{align} 
$|I_\textrm{exa}\rangle$, $|F_\textrm{exa}\rangle$, and $|a^{K=0}_\textrm{exa}\rangle$ are the exact initial, final, and intermediate states with $K=0$, respectively, and $\tau^-$ is the charge-change operator from a neutron to proton.\footnote{I have noted this operator as $\tau^+$ previously \cite{Ter12,Ter13,Ter15,Ter16}. In this paper, I change it to the convention of nuclear physics \cite{Suh07} because the GT strength functions are discussed below.} The spin-Pauli matrix is denoted by $\sigma$, and $H$ is the Hamiltonian. $M_I$ and $M_F$ are the nuclear masses of the initial and final states, respectively, and $E_{a, \textrm{exa}}^{K=0}$ is the energy of the exact intermediate state. The electron mass is denoted by $m_e c^2$. 
An abbreviation for the one-body operator
\begin{align}
\tau^- \sigma_{K=0} = \sum_{i=1}^A \tau^-(i) \sigma_{K=0}(i), 
\end{align}
is used ($i$: nucleon index). The $\mu_0$ in the left-hand side of Eq.~(\ref{eq:nme2vGT_exa}) is a sign \cite{Doi85} indicating that $M^{(2\nu)}_{GT}/\mu_0$ is dimensionless. 
In the same manner, the $2\nu\beta\beta$ Fermi NME is written 
\begin{align}
\frac{ M^{(2\nu)}_{F} }{ \mu_{0F} } =& 
\ \sum_{a^{K=0}_\textrm{exa}} \frac{1}{\mu_a} \langle F_\textrm{exa} | \tau^- | a^{K=0}_\textrm{exa}\rangle 
\langle a^{K=0}_\textrm{exa} | \tau^- | I_\textrm{exa} \rangle  \label{eq:nme2vF_exa} \\
=& \ \langle F_\textrm{exa} | \tau^- \frac{m_e c^2}{ H - \bar{M} } \tau^- | I_\textrm{exa} \rangle.  \label{eq:nme2vF_exa2}
\end{align}
The NME of the $2\nu\beta\beta$ decay is sensitive to the energy denominator $\mu_a$, thus, the closure approximation is not applied. 

Let us introduce the QRPA by replacing the nuclear states $|F_\textrm{exa}\rangle$ and $|I_\textrm{exa}\rangle$ with the corresponding QRPA states $|F\rangle$ and $|I\rangle$. The intermediate states are defined two ways; one is by the QRPA calculation using the initial ground state, and another is that using the final one. The former (latter) intermediate states are denoted by $|a^{K=0}_I\rangle$ ($|a^{K=0}_F\rangle$), with which I have two sets of equations: 
\begin{align}
\frac{ M^{(2\nu)}_{GT}(I) }{ \mu_0 } =& 
\ 3 \sum_{a^{K=0}_I, a^{K=0}_F} \langle F| \tau^- \sigma_{K=0} | a^{K=0}_F\rangle \nonumber \\
& \times \langle a^{K=0}_F | a^{K=0}_I\rangle 
\langle a^{K=0}_I | \frac{m_e c^2}{ H - \bar{M} } \tau^- \sigma_{K=0} | I \rangle,  \label{eq:nme2vGT_QRPA_I} \\
\frac{ M^{(2\nu)}_{F}(I) }{ \mu_{0F} } =& 
\ \sum_{a^{K=0}_I, a^{K=0}_F} \langle F| \tau^- | a^{K=0}_F\rangle \langle a^{K=0}_F | a^{K=0}_I\rangle \nonumber \\ 
& \times \langle a^{K=0}_I | \frac{m_e c^2}{ H - \bar{M} } \tau^- | I \rangle,  \label{eq:nme2vF_QRPA_I}
\end{align}
and
\begin{align}
\frac{ M^{(2\nu)}_{GT}(F) }{ \mu_0 } =& 
\ 3 \sum_{a^{K=0}_I, a^{K=0}_F} \langle F| \tau^- \sigma_{K=0} \frac{m_e c^2}{ H - \bar{M} } | a^{K=0}_F\rangle \nonumber \\
& \times \langle a^{K=0}_F | a^{K=0}_I\rangle \langle a^{K=0}_I | \tau^- \sigma_{K=0} | I \rangle , \label{eq:nme2vGT_QRPA_F} \\
\frac{ M^{(2\nu)}_{F}(F) }{ \mu_{0F} } =& 
\ \sum_{a^{K=0}_I, a^{K=0}_F} \langle F| \tau^- \frac{m_e c^2}{ H - \bar{M} } | a^{K=0}_F\rangle \nonumber \\
& \times \langle a^{K=0}_F | a^{K=0}_I\rangle \langle a^{K=0}_I | \tau^- | I \rangle . \label{eq:nme2vF_QRPA_F}
\end{align}
The operator $\frac{m_e c^2}{H-\bar{M}}$ includes the higher-order many quasiparticle, or many-particle-many-hole, components beyond the QRPA, thus, Eqs.~(\ref{eq:nme2vGT_QRPA_I}) and (\ref{eq:nme2vGT_QRPA_F}) [Eqs.~(\ref{eq:nme2vF_QRPA_I}) and (\ref{eq:nme2vF_QRPA_F})] do not coincide exactly. However, if the QRPA is a good approximation, the effect of those higher-order components would be small. Then, the following equations are derived: 
\begin{align}
\frac{ M^{(2\nu)}_{GT}(I) }{ \mu_0 } \simeq & 
\ 3 \sum_{a^{K=0}_I, a^{K=0}_F}  \frac{m_e c^2}{ E^{K=0}_{aI} - \bar{M} } \langle F| \tau^- \sigma_{K=0} | a^{K=0}_F\rangle \nonumber \\
& \times \langle a^{K=0}_F | a^{K=0}_I\rangle \langle a^{K=0}_I |\tau^- \sigma_{K=0} | I \rangle,  \label{eq:nme2vGT_QRPA_I2} \\
\frac{ M^{(2\nu)}_{F}(I) }{ \mu_{0F} } \simeq & 
\ \sum_{a^{K=0}_I, a^{K=0}_F}  \frac{m_e c^2}{ E^{K=0}_{aI} - \bar{M} } \langle F| \tau^- | a^{K=0}_F\rangle \nonumber \\
& \times \langle a^{K=0}_F | a^{K=0}_I\rangle \langle a^{K=0}_I |\tau^- | I \rangle,  \label{eq:nme2vF_QRPA_I2} \\
M^{(2\nu)}(I) =& \frac{ M^{(2\nu)}_{GT}(I) }{ \mu_0 } -\frac{g_V^2}{g_A^2}\frac{ M^{(2\nu)}_{F}(I) }{ \mu_{0F} } , \label{eq:nme2v_QRPA_I} \\ 
\frac{ M^{(2\nu)}_{GT}(F) }{ \mu_0 } \simeq & 
\ 3 \sum_{a^{K=0}_I, a^{K=0}_F} \frac{m_e c^2}{ E^{K=0}_{aF} - \bar{M} } \langle F| \tau^- \sigma_{K=0} | a^{K=0}_F\rangle \nonumber \\
& \times \langle a^{K=0}_F | a^{K=0}_I\rangle \langle a^{K=0}_I | \tau^- \sigma_{K=0} | I \rangle , \label{eq:nme2vGT_QRPA_F2} \\
\frac{ M^{(2\nu)}_{F}(F) }{ \mu_{0F} } \simeq & 
\ \sum_{a^{K=0}_I, a^{K=0}_F} \frac{m_e c^2}{ E^{K=0}_{aF} - \bar{M} } \langle F| \tau^- | a^{K=0}_F\rangle \nonumber \\
& \times \langle a^{K=0}_F | a^{K=0}_I\rangle \langle a^{K=0}_I | \tau^- | I \rangle , \label{eq:nme2vF_QRPA_F2} \\
M^{(2\nu)}(F) =& \frac{ M^{(2\nu)}_{GT}(F) }{ \mu_0 } -\frac{g_V^2}{g_A^2} \frac{ M^{(2\nu)}_{F}(F) }{ \mu_{0F} }, \label{eq:nme2v_QRPA_F} 
\end{align}
\begin{align}
M^{(2\nu)}(I) \simeq M^{(2\nu)}(F). \label{eq:nme2v_QRPA_FI}
\end{align}
The energy of the intermediate state is calculated using the proton-neutron (pn) QRPA excitation energies $E^{\textrm{pnQRPA}}_{K=0,aI}$ and $E^{\textrm{pnQRPA}}_{K=0,aF}$ as
\begin{align}
E^{K=0}_{aI} =& E^{\textrm{pnQRPA}}_{K=0,aI} + \lambda_p(I) - \lambda_n(I) + m_pc^2 - m_nc^2 \nonumber \\
 & + M_I, \\
E^{K=0}_{aF} =& E^{\textrm{pnQRPA}}_{K=0,aF} + \lambda_n(F) - \lambda_p(F) + m_nc^2 - m_pc^2 \nonumber \\
 &+ M_F,  
\end{align}
where $\lambda_p(I)$ and $\lambda_n(I)$ are the proton and neutron chemical potentials of the initial state, and $\lambda_p(F)$ and $\lambda_n(F)$ are those of the final state; those of the HFB ground states are used, and $m_pc^2$ and $m_nc^2$ are the proton and neutron masses. For $M_I$ and $M_F$ the experimental data are used. 
The accuracy of Eq.~(\ref{eq:nme2v_QRPA_FI}) is a consistency check point of the QRPA approach. This is shown numerically below. 

\begin{figure}[]
\includegraphics[width=1.0\columnwidth]{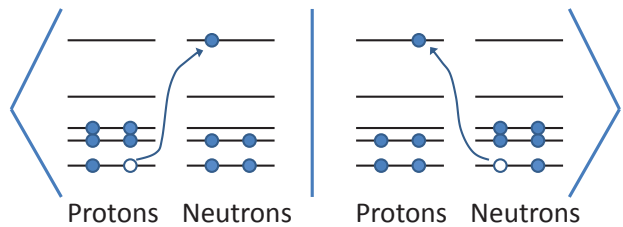}
\caption{ \label{fig:vanishing_overlap} Illustrative example of vanishing component of overlap of states obtained by charge change from $(Z+2,N-2)$- and $(Z,N)$- nuclei. }
\end{figure}

The overlap $\langle a^{K=0}_{F}| a^{K=0}_{I}\rangle$ is calculated using the equations developed in Ref.~\cite{Ter13}. However, there is a difference from the calculation of $^{150}$Nd $\rightarrow$ $^{150}$Sm \cite{Ter15,Ter16}. The norm of the unnormalized QRPA ground state, 
${\cal N}_I$ and ${\cal N}_F$ in Eqs.~(14) and (15) in Ref.~\cite{Ter13}, diverges, therefore, the norm was renormalized by truncating the contribution of the QRPA solutions so that the semiexperimental correlation energy\footnote{This is obtained from the experimental binding energy and the HFB ground-state energy.} is reproduced by the QRPA \cite{Ter15}. I used the correlation energy because this is sensitive to the QRPA correlations, and the QRPA-correlation energy diverges without the truncation \cite{Ter13}. The unnormalized overlap does not diverge because the bra and ket states are created by the charge change from the nuclei with $(\textrm{proton number}, \textrm{neutron number})=(Z+2,N-2)$ and $(Z,N)$, and many components of the unnormalized overlap vanish which keep the configuration around the Fermi surface of the HF(B) ground states (see Fig.~\ref{fig:vanishing_overlap}) \cite{Ter15}. 

Usually the Skyrme interaction (energy density functional) is constructed so as to reproduce experimental physical quantities including the binding energies of the doubly-magic nuclei by the HF ground states (Kohn-Sham states). Therefore, the HF(B) ground state replaces the (Q)RPA ground state of $^{48}$Ca in the  overlap calculation, and the HFB ground state is also used in the same manner for the ground state of $^{48}$Ti in the overlap approximately. 
In the calculation of $^{150}$Nd$\rightarrow$$^{150}$Sm, the product of the norms of the unnormalized QRPA ground states played a role to decrease the NME through a factor of $1/{\cal N}_F {\cal N}_I$, and because of this the $2\nu\beta\beta$ NME close to the semiempirical value was obtained  without very strong pn pairing interaction causing the near-instability of the QRPA solutions. Therefore it is a check point of my QRPA approach to see whether the very strong pn pairing  interaction is unnecessary for $^{48}$Ca$\rightarrow$$^{48}$Ti. 

\section{\label{sec:calculation}Calculation result}
The strength of the isovector pn [$(T,T_z)=(1,0)$] pairing interaction $g_{(T,T_z)=(1,0)}^\textrm{pair}$ is determined to be the average of the proton-proton [$(T,T_z)=(1,-1)$] and neutron-neutron [$(T,T_z))=(1,1)$] pairing interactions for satisfying approximately the isospin invariance of the $T=1$ pairing interaction. $T$ denotes the isospin, and $T_z$ is its $z$-component. 
Two QRPA calculations are performed for the $0\nu\beta\beta$ NME; one is the pnQRPA, and another is the like-particle (lp) QRPA \cite{Ter15,Ter16}. The latter can be used under the closure approximation. The calculation by the lpQRPA corresponds to the virtual decay path via the two-particle transfer. The Hamiltonian used for the HFB calculation is used for the two QRPA calculations. However, the important interaction components are different for the two QRPA calculations. Thus, the equivalence of the two paths is a theoretical constraint to the effective interactions used in the QRPA. The strength of the $T=0$ pairing interaction $g_{T=0}^\textrm{pair}$ is determined so as to have the  $0\nu\beta\beta$ GT NMEs obtained by the two methods to be equal because other interactions are established. Note that the pn pairing interactions have no contribution to the lpQRPA. 

In this paper I introduce a practical modification to my method used in Ref.~\cite{Ter16}; the Fermi component of the $0\nu\beta\beta$ NME is not used for the constraint to the effective interaction. The fundamental reason is that the effective interaction Skyrme SkM$^\ast$ plus the Coulomb interaction does not have the isospin invariance. The Fermi NME is sensitive to the $(T,T_z)=(1,0)$ pairing interaction. Thus, if the equality of the Fermi NME is required between the two different-path calculations, $g_{(T,T_z)=(1,0)}^\textrm{pair}$ can be determined. However, then, the $T=1$ pairing interaction does not satisfy the isospin invariance. The value of $g_A$ is determined so as to reproduce the experimental half-life of the $2\nu\beta\beta$ decay. Those three parameters $g_{(T,T_z)=(1,0)}^\textrm{pair}$, $g_{T=0}^\textrm{pair}$, and $g_A$ can be determined separately in this order. 
I also use this $g_A$ to the NME calculation of $0\nu\beta\beta$ decay because the very large single-particle space is used. 
If this space is not enough large, different $g_A$'s would be necessary for the two decays because the neutrino potential of the $0\nu\beta\beta$ decay has a divergence. 
The value of $g_V$ is always 1 throughout this paper. 

The value of $g_{(T,T_z)=(1,0)}^\textrm{pair}$ is $-241.43$ MeV$\,$fm$^3$, and $g_{T=0}^\textrm{pair}$ was found to be $-180.0$ MeV$\,$fm$^3$ according to the above method. 
The experimental half-life of the $2\nu\beta\beta$ decay of $^{48}$Ca is $T_{1/2}^{(2\nu)\textrm{exp}}$ = (6.4$\pm$1.2)$\times$10$^{19}$ yr \cite{Arn16}. 
The corresponding theoretical half-life is calculated by \cite{Kot12}  
\begin{align}
T_{1/2}^{(2\nu)\textrm{th}} =&  ( G_{2\nu}^{(0)} g_A^4 )^{-1} \left| M^{(2\nu)} \right|^{-2},  
\end{align}
with the phase-space factor $G_{2\nu}^{(0)}$ = 15550$\times$$10^{-21}$ yr$^{-1}$ \cite{Kot12}. 
$T_{1/2}^{(2\nu)\textrm{exp}}$ was reproduced by 
$g_A$ = 0.48 with $M^{(2\nu)}(I)$ = 0.138 and 
$g_A$ = 0.49 with $M^{(2\nu)}(F)$ = 0.133. 
The relative difference of $M^{(2\nu)}(I)$ and $M^{(2\nu)}(F)$ is $\simeq$ 4 \%, thus  the consistency of the QRPA approach discussed above is approximately satisfied. The GT and Fermi NMEs are shown in Table \ref{tab:2vbbNME}. The absolute value of $M^{(2\nu)}_F/\mu_{0F}$ is less than 5\% of $M^{(2\nu)}_{GT}/\mu_0$, thus, the isospin invariance of $M^{(2\nu)}$ is also approximately satisfied. 
\begin{table}
\caption{\label{tab:2vbbNME} Calculated values of $2\nu\beta\beta$ NMEs and $T_{1/2}^{(2\nu)}$ for $^{48}$Ca$\rightarrow$$^{48}$Ti and $g_A$. }
\begin{ruledtabular}
\begin{tabular}{cccccc}
Equations & $g_A$ & $M^{(2\nu)}$ & $\frac{M^{(2\nu)}_{GT}}{\mu_0}$ & $\frac{M^{(2\nu)}_{F}}{\mu_{0F}}$ & $T_{1/2}^{(2\nu)}$ ($10^{19}$yr) \\
\hline
(\ref{eq:nme2vGT_QRPA_I2})$-$(\ref{eq:nme2v_QRPA_I}) & 0.48 & 0.138 & 0.124 & $-0.0033$ & 6.339 \\
(\ref{eq:nme2vGT_QRPA_F2})$-$(\ref{eq:nme2v_QRPA_F}) & 0.49 & 0.133 & 0.112 & $-0.0052$ & 6.274 
\end{tabular}
\end{ruledtabular}
\end{table}

By using the result of Ref.~\cite{Ter15} for $^{150}$Nd, it turns out that $g_A=0.84$ reproduces 
$T_{1/2}^{(2\nu)\textrm{exp}}=8.2$$\times$$10^{18}$ yr \cite{Bar13} of that nucleus. The $g_A$ value of $^{48}$Ca  is 58\% of that of $^{150}$Nd. One of the causes for this difference is apparently the normalization factors of the QRPA ground states. 
The product of the two normalization factors of $^{150}$Nd and $^{150}$Sm is 1.860 \cite{Ter15}, but the corresponding product of 
$^{48}$Ca and $^{48}$Ti is 1.0; see the previous section. If the product of the normalization factors of 1.860 is applied artificially to the $^{48}$Ca calculation, the $T^{(2\nu)\textrm{exp}}_{1/2}$ is reproduced by $g_A=0.68$. This value is larger than 0.49 as expected, although still smaller than 0.84 of $^{150}$Nd.  Other reasons may be the differences in the nuclear structure of those nuclei, however they are not obvious. 
The $g_A$ of my calculation is consistent with a recent tendency to accept $g_A < 1.0$, even $g_A \approx 0.5$ (not for $^{48}$Ca), e.g., \cite{Suh17}. For the $0\nu\beta\beta$ decays of $^{48}$Ca, the bare value  of 1.25 or 1.27 is usually used by other groups; see the comparison below. 

\begin{figure}[t]
\includegraphics[width=\columnwidth]{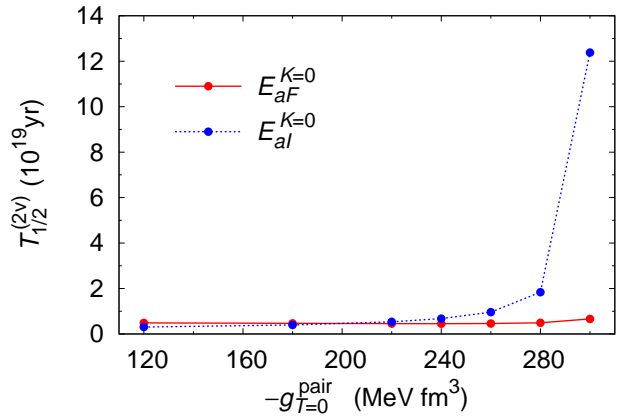}
\caption{ \protect \label{fig:t0_halflife_ga1p0} Calculated half-life of $^{48}$Ca to $2\nu\beta\beta$ decay as functions of $-g^\textrm{pair}_{T=0}$. 
The points connected by solid (dotted) lines were obtained using $E^{K=0}_{aF}$'s ($E^{K=0}_{aI}$'s). 
The parameters $g_A=1.0$ and $g^\textrm{pair}_{(T,T_z)=(1,0)}=-241.43$ MeV$\,$fm$^3$ were used. }
\end{figure}

I performed a reference calculation using a usual method; $g^\textrm{pair}_{T=0}$ was 
determined so as to reproduce the $T^{(2\nu)\textrm{exp}}_{1/2}$. For $g_A$, an effective value of 1.0 was used, and I used $g^\textrm{pair}_{(T,T_z)=(1,0)}$ already determined. The $g^\textrm{pair}_{T=0}$-dependence of 
$T^{(2\nu)}_{1/2}$ is shown in Fig.~\ref{fig:t0_halflife_ga1p0}. The difference in input for the two sets of result is whether $E_{aF}^{K=0}$'s are used (connected by solid line) or $E_{aI}^{K=0}$'s are used (connected by dotted line). The mean value of the two results at $g^\textrm{pair}_{T=0}=-300.0$ MeV$\,$fm$^3$ (the rightmost points) is close to the experimental value of (6.4$\pm 1.2)$$\times 10^{19}$ yr. However, the discrepancy of the two points is too large, therefore the QRPA is not a good approximation. This method to determine $g^\textrm{pair}_{T=0}$ cannot be used. 
Generally, the instability of the mean-field or HFB ground state occurs in relation to the symmetry breaking, if the strength of attractive interaction increases significantly. The QRPA is usually not used near this instability. 

\begin{table*}
\caption{\label{tab:comp_M0v_all} $M^{(0\nu)}_{GT}$, $M^{(0\nu)}_{F}$, $M^{(0\nu)}_{T}$, $M^{(0\nu)}$, $g_A$, $R^{(0\nu)}_{1/2}$, and $M^{(0\nu)\prime}$ of the $0\nu\beta\beta$ decay of $^{48}$Ca calculated by different groups. The first column indicates the group. 
The Calculation 2 is my calculation. The Calculations 1 and 3$-$10 indicate the results of Refs.~\cite{Sim13} (Argonne V18), \cite{Men09}, \cite{Sen13}, \cite{Iwa16}, \cite{Bar15}, \cite{Vaq13}, \cite{Yao15}, \cite{Jia17}, and \cite{Kwi14}, respectively. 
Three results are shown in Calculation 1. The upper two rows show the results with no pairing gap of $^{48}$Ca ground state, and two values of $g_A$ are used: 1.0 for the first row and 1.27 for the second row. 
The third row shows the result with finite pairing gaps of the ground states and $g_A=1.27$, see Ref.~\cite{Sim13} for the pairing gaps. 
The difference in the two results of Calculation 3 is in the method to modify the neutrino potential in terms of the short-range correlations: so-called the SRC.  
The two values of Calculation 5 correspond to the minimum and maximum $M^{(0\nu)}$ obtained using the two major shells. 
Some effective method of the SRC is used except for Calculations 2, 8, and 9.   For Calculation 10, see Ref.~\cite{Kwi14}. 
The mark of $\ast$ indicates that the specified term is included in the calculation, however, the value is not noted in the paper. 
The double-$\ast$ mark indicates that the term is not included in the calculation. 
The definition of the sign of $M^{(0\nu)}_T$ is unified to that of Ref.~\cite{Sim13}. 
$M^{(0\nu)}_T$ of Calculation 8 is not obvious because of the different theoretical framework. 
In Calculation 10 \cite{Kwi14}, $(g_V^2/g_A^2)M^{(0\nu)}_F = -0.160$ is obtained, but $g_A$ is not noted in the paper. 
See also Ref.~\cite{Suh93}.}
\begin{ruledtabular}
\begin{tabular}{clllllll}
 Cal. & $M^{(0\nu)}_{GT}$ & \ $M^{(0\nu)}_F$ & \ $M^{(0\nu)}_T$ & $M^{(0\nu)}$ & \ \ $g_A$ & 
$R^{(0\nu)}_{1/2}$  & $M^{(0\nu)\prime}$ \\
 & & & & & & \!\!\!\!\!\!\!\!\!\!\!($10^{12}$\,MeV$^2$\,yr) & \\
\hline
\multirow{3}{*}{\ \ \ 1$\left\{\vbox{\vspace{20.5pt}}\right.$} & 0.639 & $-$0.268 & $-$0.161 & 0.745 & 1.0 & 18.95 & 0.462 \\
 & 0.523 & $-$0.268 & $-$0.149 & 0.541 & 1.27 & 13.82 & 0.541 \\
 &    \ \ \ $\ast$      &   \ \ \ \ $\ast$            &     \ \ \ \ $\ast$          & 0.71    & 1.27 & \,\,\,8.02 & 0.71 \\
2 & 1.723 & $-$0.319 & \ \ \ $\ast\ast$ & 3.054 & 0.49 & 19.572 & 0.454\\
\multirow{2}{*}{\ \ \,3$\left\{\vbox{\vspace{14pt}}\right.$} & 0.575 & $-$0.144 & $-$0.057 & 0.61 & 1.25 & 11.585 & 0.591 \\
 & \ \ \ $\ast$ & \ \ \ \ $\ast$ & \ \ \ \ $\ast$ & 0.85 & 1.25 & \,\,\,5.966 & 0.823 \\
4 & 0.747 & $-$0.208 & $-$0.079 & 0.800 & 1.254 & \,\,\,6.650 & 0.780 \\
\multirow{2}{*}{\ \ \,5$\left\{\vbox{\vspace{14pt}}\right.$} & 0.852 & $-$0.288 & $-$0.068 & 0.963 & 1.27 & \,\,\,4.389 & 0.963 \\
 & 1.045 & $-$0.327 & $-$0.065 & 1.183 & 1.27 & \,\,\,2.905 & 1.183 \\
6 & 1.73 & $-$0.30 & $-$0.17 & 1.75 & 1.269 & \,\,\,1.325 & 1.75\\
7 & 1.793 & $-$0.673 & \ \ \ $\ast\ast$ & 2.229 & 1.25 & \,\,\,0.867 & 2.16 \\
8 & \ \ \ $\ast$ & \ \ \ \ $\ast$ & & 3.66 & 1.254 & \,\,\,0.317 & 3.57 \\
9 & \ \ \ $\ast$ & \ \ \ \ $\ast$ & \ \ \ \ $\ast$ & 1.082 & \!\!\!\!$\approx$1.27 & \,\,\,3.455 & 1.082 \\
10 & 1.211 & \ \ \ \ $\ast$ & $-$0.070 & 1.301 & \ \ \ $\ast$ & & 
\end{tabular}
\end{ruledtabular}
\end{table*}

Subsequently, I calculated the $0\nu\beta\beta$ NME $M^{(0\nu)}$ (for the equation, see Ref.~\cite{Ter15}) using $g_A=0.49$, and 
$M^{(0\nu)}$ = 3.054 was obtained. My effective $g_A$ is much smaller than the usual ones $\sim$ 1.0. 
In Table \ref{tab:comp_M0v_all}, $M^{(0\nu)}$, $M^{(0\nu)}_{GT}$ (Gamow-Teller $0\nu\beta\beta$ NME), $M^{(0\nu)}_F$ (Fermi $0\nu\beta\beta$ NME), $M^{(0\nu)}_T$ ($0\nu\beta\beta$ NME of the tensor transition operator, shown if used), and $g_A$ of the different groups are shown. 
For comparison of the results with different $g_A$'s, I also show in the table the reduced half-life
\begin{align}
R_{1/2}^{(0\nu)} = (G_{0\nu} g_A^4 )^{-1} (m_e c^2)^2 | M^{(0\nu)}|^{-2}, \label{eq:reduced_halflife_0vbb}
\end{align}
where $G_{0\nu}$ is the phase-space factor of the $0\nu\beta\beta$ decay 
(in my calculation, $G_{0\nu}$ = 0.2481$\times$10$^{-13}$ yr$^{-1}$ \cite{Kot12}), and the scaled $0\nu\beta\beta$ NME, e.g.~\cite{Sim13}, \begin{align}
M^{(0\nu)\prime} = \frac{ g_A^2 }{ (g_A^\textrm{bare})^2 } M^{(0\nu)},
\end{align}
where $g_A^\textrm{bare}$ is the value of $g_A$ not including the many-body effect or compensation of approximation, and $g_A^\textrm{bare}$ = 1.27 is used, e.g., \cite{Eng17}. 
The methods of Calculations 1 and 2 are the (Q)RPA (the latter is my calculation), those of 3$-$5 are the (interacting) shell-model, that of 6 is the interacting-boson model, and those of 7$-$9 are the generator-coordinate method, see the caption for the references. 
$R_{1/2}^{(0\nu)}$ is the quantity used for deriving the effective neutrino mass $\langle m_\nu \rangle$ in 
\begin{align}
\langle m_\nu \rangle^2 = \frac{ R_{1/2}^{(0\nu)} }{ T_{1/2}^{(0\nu)} }, \label{eq:effective_v_mass}
\end{align}
with the half-life of the $0\nu\beta\beta$ decay $T_{1/2}^{(0\nu)}$ expected to be available experimentally. 
$T_{1/2}^{(0\nu)}$ and $\langle m_\nu \rangle$ are unique; if all calculations using different approximations are correct, $R_{1/2}^{(0\nu)}$ would be identical. Therefore $R^{(0\nu)}_{1/2}$ is better than $M^{(0\nu)}$ for comparison of different calculations. 
$R^{(0\nu)}_{1/2}$ of my calculation, Calculation 2, is close to the largest one of another (Q)RPA calculation. The QRPA calculations show the largest $R_{1/2}^{(0\nu)}$ in the calculations by the different methods. The $T_{1/2}^{(0\nu)}$ predicted by my calculation is larger than that by the shell model \cite{Iwa16} by a factor of five. 

It is noted that I used the latest  value of $T_{1/2}^{(2\nu)\textrm{exp}}$, which is nearly 50 \% larger than the previous values \cite{Bar13}. 
However, the old value is used in some of the other calculations. If the old one, 4.4$\times$10$^{19}$ yr, is used for the fitting, my $R_{1/2}^{(0\nu)}$ is 
16.297$\times$10$^{12}$ MeV$^2$yr [$g_A$ = 0.53, Eqs.~(\ref{eq:nme2vGT_QRPA_I2})$-$(\ref{eq:nme2v_QRPA_I})] and 
15.574$\times$10$^{12}$ MeV$^2$yr [$g_A$ = 0.54, Eqs.~(\ref{eq:nme2vGT_QRPA_F2})$-$(\ref{eq:nme2v_QRPA_F})]. Thus, there is no qualitative influence on the comparison. 

As seen from Eqs.~(\ref{eq:reduced_halflife_0vbb})$-$(\ref{eq:effective_v_mass}), $M^{(0\nu)\prime}$ is also a quantity used for obtaining $\langle m_\nu \rangle$. Other necessary input for obtaining $\langle m_\nu\rangle$ are the experimental data and constants other than $g_A$, thus, $M^{(0\nu)\prime}$ can also be compared between calculations with different $g_A$. This NME is less affected by the uncertainty of $g_A$ than $R_{1/2}^{(0\nu)}$ because the $g_A$ dependence of $M^{(0\nu)\prime}$ is not higher order than  $g_A^2$. The most significant difference between $R^{(0\nu)}_{1/2}$ and $M^{(0\nu)\prime}$ is seen in Calculations 1$-$4. The results of these four calculations seem close in  $M^{(0\nu)\prime}$, but the differences in $R^{(0\nu)}_{1/2}$ are significantly larger. Difference as this can occur because of the relation 
\begin{align}
R_{1/2}^{(0\nu)} \propto (M^{(0\nu)\prime})^{-2}. 
\end{align}
When $M^{(0\nu)\prime}$'s are $\simeq$ 0.5 or smaller than this, the difference between them is magnified in $R_{1/2}^{(0\nu)}$. 
Both $R_{1/2}^{(0\nu)}$ and $M^{(0\nu)\prime}$ are important. 
One can concentrate on the nuclear property in discussing $M^{(0\nu)\prime}$. 
On the other hand, the half-life is the physical quantity measured, and $R_{1/2}^{(0\nu)}$ is more directly related to that than $M^{(0\nu)\prime}$ because of the proportionality. 

The bare value of $g_A$ is usually used in the $^{48}$Ca calculations. Note, however, that in the larger set of samples of other nuclei including the single-$\beta$ decays, the effective value of 1.0 is historically more usual, e.g.,~\cite{Bro85}. My $M^{(0\nu)}_{GT}$ and $M^{(0\nu)}_F$ are close to those of Calculation 6 (the interacting-boson model), however $M^{(0\nu)}$ are quite different because of the difference in the used $g_A$. In my calculation, that $g_A$ is necessary for reproducing the $T_{1/2}^{(2\nu)\textrm{exp}}$. It is seen from the comparison of the first two rows of Calculations 1 and 2 (both are the QRPA calculations) that $M^{(0\nu)}$ and $g_A$ are quite different, however, $R^{(0\nu)}_{1/2}$ and $M^{(0\nu)\prime}$ are close. 
Both calculations use $T_{1/2}^{(2\nu)\textrm{exp}}$ for determining parameter, but the methods to determine $g^\textrm{pair}_{T=0}$ and $g_A$ are different. 
The fluctuation of $M^{(0\nu)}_{GT}$ seen in Table \ref{tab:comp_M0v_all} is larger than those of $M^{(0\nu)}_F$ and $M^{(0\nu)}_T$. 

For a reference, I also made a comparison under as unified conditions as possible; $R^{(0\nu)}_{1/2}(g_A=1.0)$ and $M^{(0\nu)\prime}(g_A=1.0)$ calculated with the same $g_A$ = 1.0  and without $M^{(0\nu)}_T$ are shown in Table \ref{tab:comp_same_gA} for those calculations in which $M^{(0\nu)}_{GT}$ and $M^{(0\nu)}_{F}$ are available. The method dependence of $M^{(0\nu)\prime}(g_A=1.0)$ is similar to that of $M^{(0\nu)}_{GT}$, as seen from the comparison of my result and the others in Tables \ref{tab:comp_M0v_all} and \ref{tab:comp_same_gA}.  
An analogous tendency (but the inverted method dependence) is seen for $R^{(0\nu)}_{1/2}(g_A=1.0)$. 
My values of $M^{(0\nu)\prime}$ and $R^{(0\nu)}_{1/2}$ change much more than those of other groups between the two tables because my value of $g_A=0.49$ is much smaller than those used by other groups. 
\begin{table}
\caption{\label{tab:comp_same_gA}$R^{(0\nu)}_{1/2}(g_A=1.0)$  and $M^{(0\nu)\prime}(g_A=1.0)$ calculated with $g_A=1.0$ for  Calculations 1$-$7 without $M^{(0\nu)}_T$. The calculation number and constituent results correspond to those rows in Table \ref{tab:comp_M0v_all}. Mark of  $\ast$ indicates that $M^{(0\nu)}_{GT}$ and $M^{(0\nu)}_F$ are not available.}
\begin{ruledtabular}
\begin{tabular}{crr}
 Cal. &  $R^{(0\nu)}_{1/2}(g_A=1.0)$  & $M^{(0\nu)\prime}(g_A=1.0)$ \\
 &  ($10^{12}$\,MeV$^2$\,yr) & \\
\hline
\multirow{3}{*}{\ \ \ 1$\left\{\vbox{\vspace{21pt}}\right.$} & 12.79\hbox{\hspace{25pt}}  & 0.562\hbox{\hspace{25pt}}  \\
 & 16.82\hbox{\hspace{25pt}} &  0.490\hbox{\hspace{25pt}}  \\
 &  $\ast$\hbox{\hspace{32pt}} &  $\ast$\hbox{\hspace{32pt}}  \\
2 & 2.52\hbox{\hspace{25pt}}  &  1.266\hbox{\hspace{25pt}} \\
\multirow{2}{*}{\ \ \,3$\left\{\vbox{\vspace{14pt}}\right.$} & 20.36\hbox{\hspace{25pt}} & 0.446\hbox{\hspace{25pt}} \\
 & $\ast$\hbox{\hspace{32pt}} & $\ast$\hbox{\hspace{32pt}} \\
4 &  11.54\hbox{\hspace{25pt}} & 0.592\hbox{\hspace{25pt}} \\
\multirow{2}{*}{\ \ \,5$\left\{\vbox{\vspace{14pt}}\right.$} & 8.10\hbox{\hspace{25pt}}  &  0.707\hbox{\hspace{25pt}} \\
 &  5.52\hbox{\hspace{25pt}} &  0.856\hbox{\hspace{25pt}} \\
6 & 2.55\hbox{\hspace{25pt}}  & 1.259\hbox{\hspace{25pt}} \\
7 &  1.73\hbox{\hspace{25pt}} & 1.529\hbox{\hspace{25pt}} 
\end{tabular}
\end{ruledtabular}
\end{table}

\begin{figure}[]
\includegraphics[width=1.0\columnwidth]{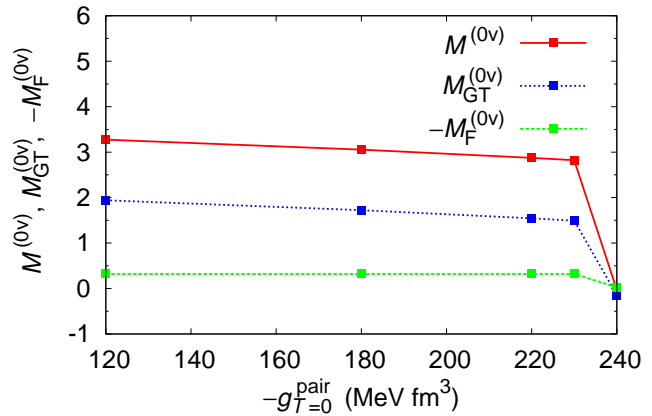}
\caption{ \label{fig:t0_nme} Dependence of $M^{(0\nu)}$, $M^{(0\nu)}_{GT}$, and $-M^{(0\nu)}_F$  on $g_{T=0}^\textrm{pair}$.}
\end{figure}

A possible difference between the QRPA calculation of Ref.~\cite{Sim13} and mine is the $T=0$ pairing-interaction strength because  my original method is used for determining that strength. Since the different operators are used for defining that interaction, the interaction strength can only be compared in terms of the position on the curve of the NME versus $g^\textrm{pair}_{T=0}$ (this information of Ref.~\cite{Sim13} is not available). 
I show in Fig.~\ref{fig:t0_nme} the plots of $M^{(0\nu)}$, $M^{(0\nu)}_{GT}$, and $-M^{(0\nu)}_F$ versus $-g^\textrm{pair}_{T=0}$ of my calculation. The adopted value of $g_{T=0}^\textrm{pair}$ is $-$180.0 MeV\,fm$^3$, as mentioned above, thus, it is seen that my calculation is in the safe region of the QRPA. This is reasonable because $g_{T=0}^\textrm{pair}$ is determined referring to $M^{(0\nu)}_{GT}$ obtained by the lpQRPA \cite{Ter16}, for which the HFB ground state is stable. The specialty of the overlap of $^{48}$Ca does not cause a problem. 

%

\section{\label{sec:GT_strength_function} Gamow-Teller strength function}
\subsection{\label{sec:brief_review}Brief review}
The GT strength function is calculated using the GT transition matrix, of which transition density is also an ingredient of the NMEs of the $\beta\beta$ decay. Naturally the GT strength function obtained from the experiments of the $(p,n)$ and $(n,p)$ reactions are important information for checking a part of the calculated $\beta\beta$ NMEs. This strength function and the half-life of the 2$\nu\beta\beta$ decay are the most direct experimental data helping the calculation of the  $0\nu\beta\beta$ NME because the $\beta$ decay of $^{48}$Ca is  suppressed by the very small Q value of 279 keV \cite{nndc}, and $^{48}$Ti does not have the $\beta^+$ decay or electron capture. 

The above charge exchange occurs in the hadron knock-on reaction, therefore, the mechanism is independent of $g_A$. Thus, the charge-exchange reaction seems to be a quite adequate method for checking the GT transition matrix elements. However, the extraction of the GT strength function from the experimental cross section is not straightforward, in addition, there was a historical problem of the quenching of the measured GT strength. 
I briefly review those discussions. 

The excitation by the spin-isospin operator of $\sigma\tau$ is effectively induced by the $(p,n)$ and $(n,p)$ reactions with the incident energy of 200 MeV or larger at the forward angles. The basic equation for extracting the GT transition strength $B$(GT) from the cross section $\sigma_\textrm{GT}(q,\omega)$ is  
\begin{align}
\sigma_\textrm{GT}(q,\omega) = \hat{\sigma}_\textrm{GT} F(q,\omega) B(\textrm{GT}), \label{eq:proportionality_eq}
\end{align}
where $\hat{\sigma}_\textrm{GT}$ is the unit cross section determined experimentally, and $F(q,\omega)$ is a function depending on the momentum transfer $q$ and energy loss $\omega$ (variables) , see, e.g., Ref.~\cite{Wak97} for these factors. For the derivation of Eq.~(\ref{eq:proportionality_eq}), sometimes called the proportionality relation, see, e.g., Ref.~\cite{Ich06}. 
The limit of $q \rightarrow 0$ is used for extracting $B(GT)$ from the exeprimental cross section \cite{Wak97,Ich06}. 
The presumption for Eq.~(\ref{eq:proportionality_eq}) is that the reaction is induced by a one-body field (the impulse approximation) \cite{Ost85, Tad87} and a single-step reaction \cite{Wat93}. 
There is also another method for determining $B(\textrm{GT})$ by Eq.~(\ref{eq:proportionality_eq}) and the experimental data of the $\beta$-decay for pairs of mirror nuclei \cite{Mol15} (not applied to $^{48}$Ca and $^{48}$Ti).

The quenching factor of the sum of the experimental charge-exchange strengths corresponding to the GT sum rule (Ikeda sum rule) \cite{Ike63} is defined by  
\begin{align}
Q = \frac{ \mathcal{S}_{\beta^-}^\textrm{exp} - \mathcal{S}_{\beta^+}^\textrm{exp} }{3(N-Z)}, 
\end{align}
where $\mathcal{S}_{\beta^-}^\textrm{exp}$ and $\mathcal{S}_{\beta^+}^\textrm{exp}$ are the sums of the experimental transition strengths of the $\beta^-$- and $\beta^+$- decay type, respectively. In the early days, this $Q$ was 0.40$-$0.65 systematically in a broad mass region \cite{Ich06}. 
This problem stimulated the discussion on the contribution of the $\Delta$-isobar nucleon hole; for this see the references in, e.g., Ref.~\cite{Ich06}. 
Below is the history of the studies on the basis of the nucleon degrees of freedom. 

The cross sections and deduced strength functions 
are reported by several experiments for mother nuclei $^{90}$Zr \cite{Yak05, Pro00,Wak97,Ray90} and $^{208}$Pb \cite{Pro00,Ray90,Gaa81}. The strength function of $^{90}$Zr({\it p},{\it n}) \cite{Yak05} consists of a sharp peak around $E=1$ MeV, the giant resonance in $E=5$$-$20 MeV, and a broad and low strength distribution in $E \geq 20$ MeV.  
The transition strength of $^{90}$Zr({\it n},{\it p}) is much smaller, as anticipated for the neutron-excess nucleus, but has a non-negligible broad distribution. 
That of $^{208}$Pb({\it p},{\it n}) has a similar structure of the giant resonance and broad distribution. 

It has been found \cite{Pro00} that the broad and low distribution in the high-energy region was seen with the incident energy of 795 MeV but not seen with 200 MeV. The authors of that paper argue that the projectile with the higher energy can be absorbed more efficiently than that with the lower energy, thus the high-energy broad distribution is due to the isovector spin monopole excitation. That is a compression mode and induced by the transition operator $r^2\sigma \tau$. Theoretically, this possibility was discussed in, e.g., 
Ref.~\cite{Aue89}. 

The authors of Refs.~\cite{Dan97} and \cite{Ber82} showed independently that the distribution of the transition strength was shifted substantially to the high-energy side by the 2-particle-2-hole (2p-2h) correlations of the nuclear states. It is noted that the charge-exchange transition strengths due to the above two mechanisms have similar high-energy broad distributions if scaled to the same height.  
It has been pointed out \cite{Ham00} that $r^2\sigma \tau$ induces not only the compression mode but also the GT transition, and a method to separate these two components was suggested. 

The experimental GT strengths in which the isovector spin monopole component has been subtracted were derived \cite{Yak05} for $^{90}$Zr({\it n},{\it p}) and $^{90}$Zr({\it p},{\it n}) reactions up to $E=70$ MeV, and the quenching factor was found to be $Q=0.88\pm0.06$. This is a value much close to the unity compared to those in the early days. Summarizing the status of the $^{90}$Zr studies, the quenching problem of the experimental GT strength seems solvable by extending the measured energy region. The high-energy ($E \geq 20$ MeV) broad and low distribution of the strength in the original data contains both the 2p-2h and the isovector spin monopole excitations (this is 1p-1h excitations). 

\subsection{\label{sec:GT_strength_function_48Ca} Gamow-Teller strength functions of $^{\bm{48}}$Ca and $^{\bm{48}}$Ti}
Now I examine my transition strengths of $^{48}$Ca and $^{48}$Ti. 
I obtained the calculated value of the GT sum rule of 
\begin{align}
{\cal S}_{\beta^-}^\textrm{QRPA} - {\cal S}_{\beta^+}^\textrm{QRPA} &=24.638 - 0.633 = 24.005, \ (^{48}\textrm{Ca}), \nonumber \\ 
{\cal S}_{\beta^-}^\textrm{QRPA} - {\cal S}_{\beta^+}^\textrm{QRPA} &=15.257 - 3.268 = 11.989, \ (^{48}\textrm{Ti}). 
\nonumber
\end{align}
$\mathcal{S}_{\beta^-}^\textrm{QRPA}$ and $\mathcal{S}_{\beta^+}^\textrm{QRPA}$ are the GT strength sums calculated by the QRPA corresponding to $\mathcal{S}_{\beta^-}^\textrm{exp}$ and $\mathcal{S}_{\beta^+}^\textrm{exp}$, respectively. 
The exact sum-rule values are 24 ($^{48}$Ca) and 12 ($^{48}$Ti). Thus the QRPA calculation satisfies the sum rule accurately; this is a technical check of the QRPA calculation. 

\begin{figure*}[]
\centering
\begin{minipage}[t]{.48\textwidth}
\includegraphics[width=1.0\columnwidth]{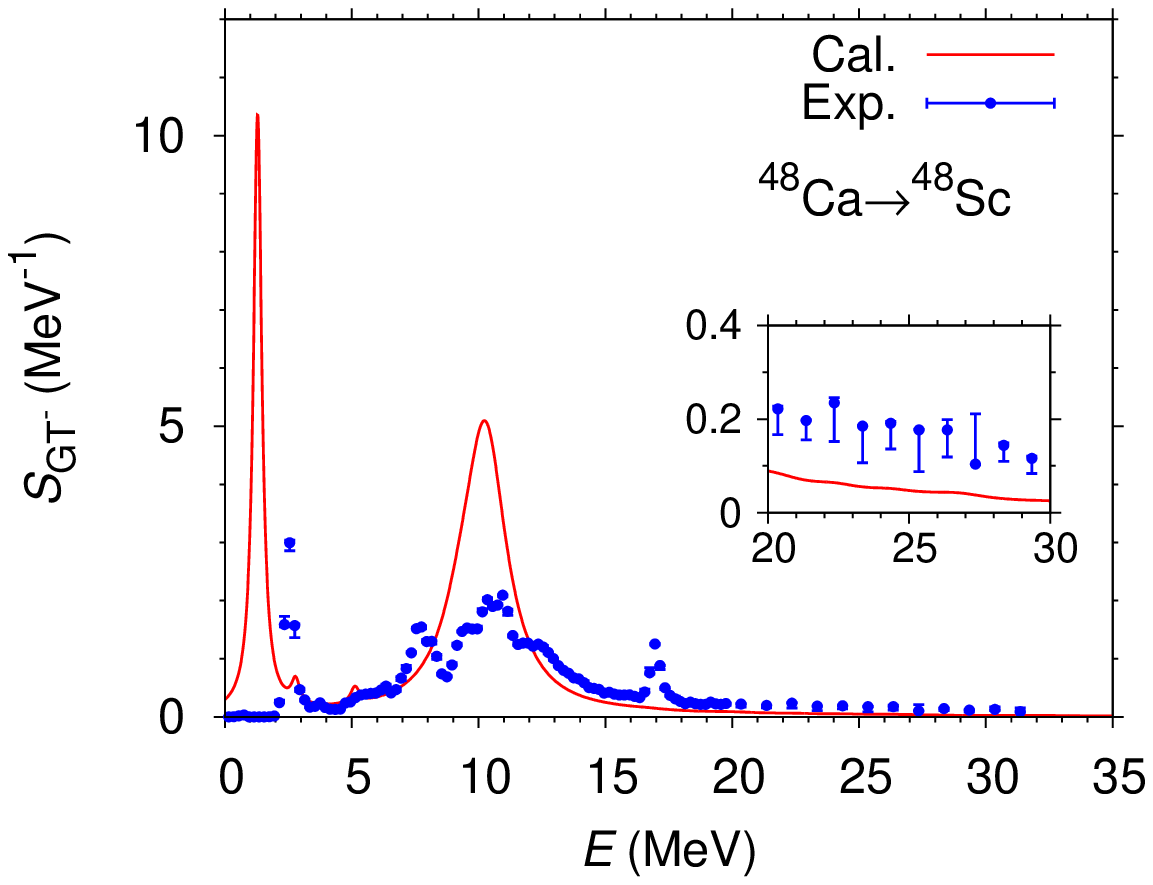}
\caption{ \label{fig:strfn_gt_48Ca} Strength functions of the GT transition from $^{48}$Ca to $^{48}$Sc measured (isolated points with error bars) and calculated by the QRPA (solid line). The origin of the excitation energy $E$ is at the ground state of $^{48}$Sc. The measured one \cite{Yak09} is obtained using Eq.~(\ref{eq:proportionality_eq}) [$dB(GT)/dE$]. The inset is a magnification of the high-energy region.}
\end{minipage}
\hspace{15pt}
\begin{minipage}[t]{.47\textwidth}
\includegraphics[width=1.0\columnwidth]{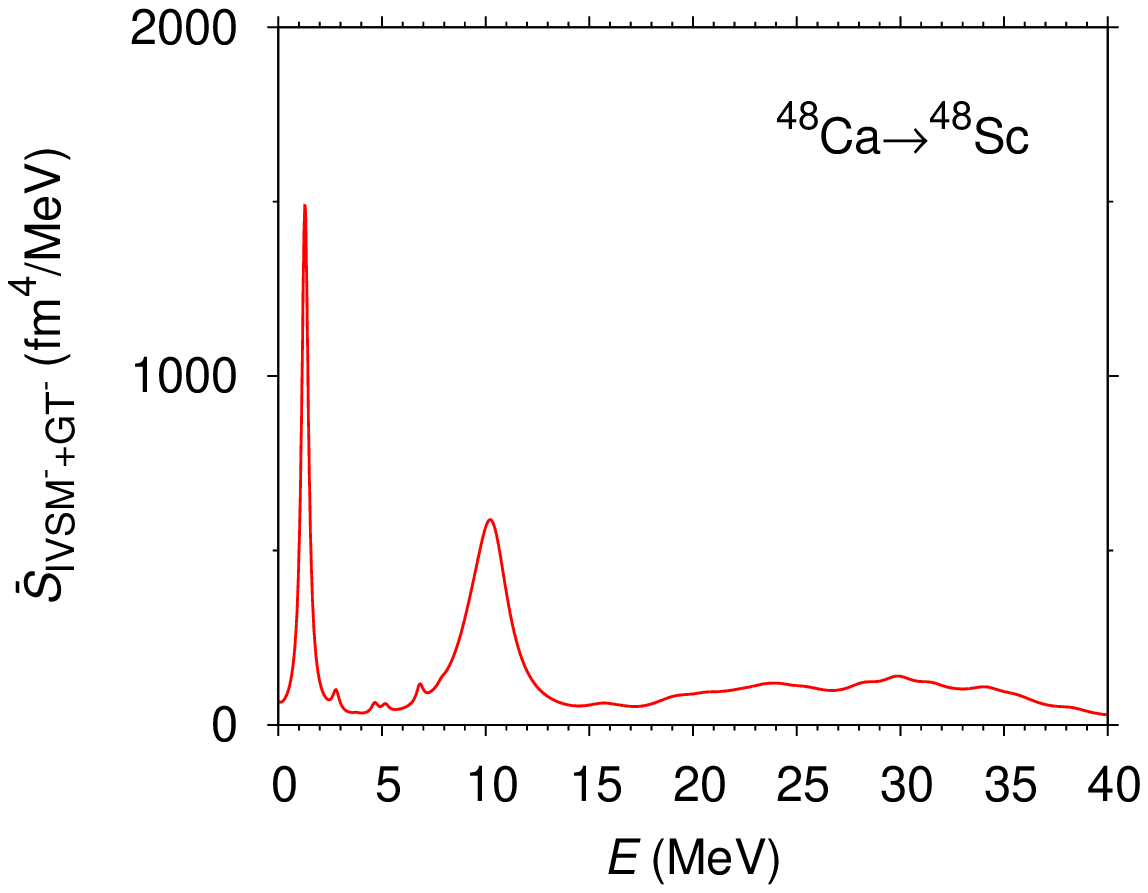}
\caption{ \label{fig:strfn_ivsm_48Ca} The same as the solid line in Fig.~\ref{fig:strfn_gt_48Ca} but for the transition operator $r^2\sigma\tau$.}
\end{minipage}
\end{figure*}

\begin{figure*}[]
\begin{minipage}[t]{.47\textwidth}
\includegraphics[width=1.0\columnwidth]{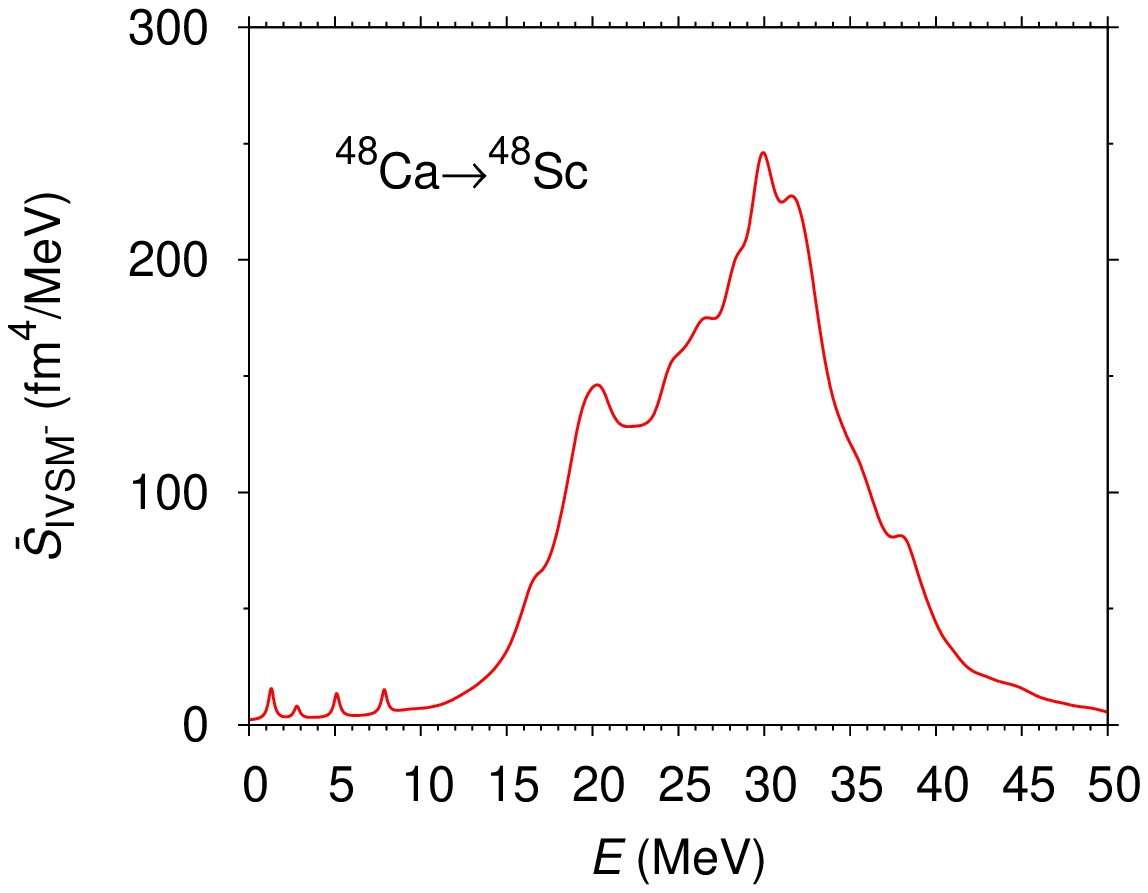}
\caption{ \label{fig:strfn_gt-sbtd_ivsm_48Ca} The same as Fig.~\ref{fig:strfn_ivsm_48Ca} but for the transition operator $(r^2-\langle r^2 \rangle_\textrm{n1f7/2})\sigma\tau$.}
\end{minipage}
\hspace{15pt}
\begin{minipage}[t]{.47\textwidth}
\includegraphics[width=1.0\columnwidth]{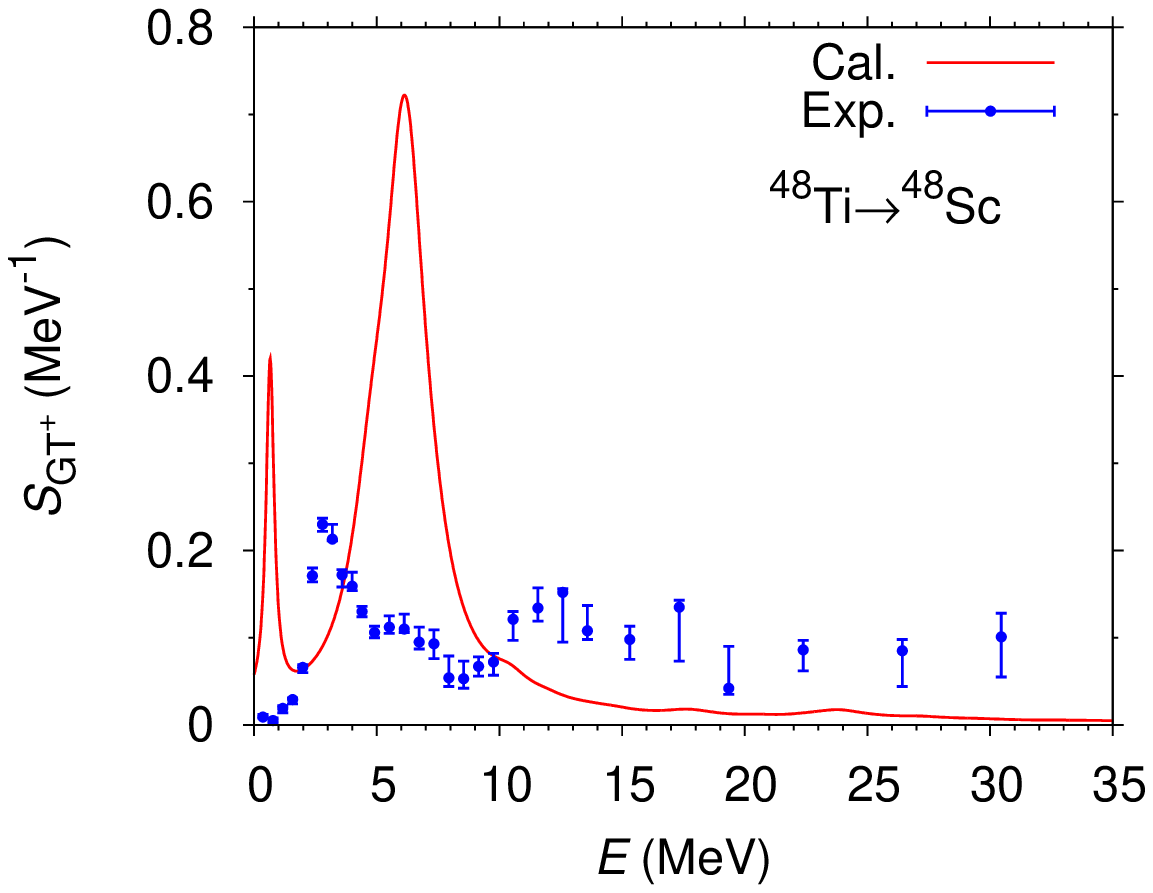}
\caption{ \label{fig:strfn_gt_48Ti} The same as Fig.~\ref{fig:strfn_gt_48Ca} but for $^{48}$Ti $\rightarrow$ $^{48}$Sc. The width parameter of 1.0 MeV is used for the excitation energies larger than 4 MeV for simulating the experimental width.}
\end{minipage}
\end{figure*}

\begin{figure*}[]
\begin{minipage}[t]{.47\textwidth}
\includegraphics[width=1.0\columnwidth]{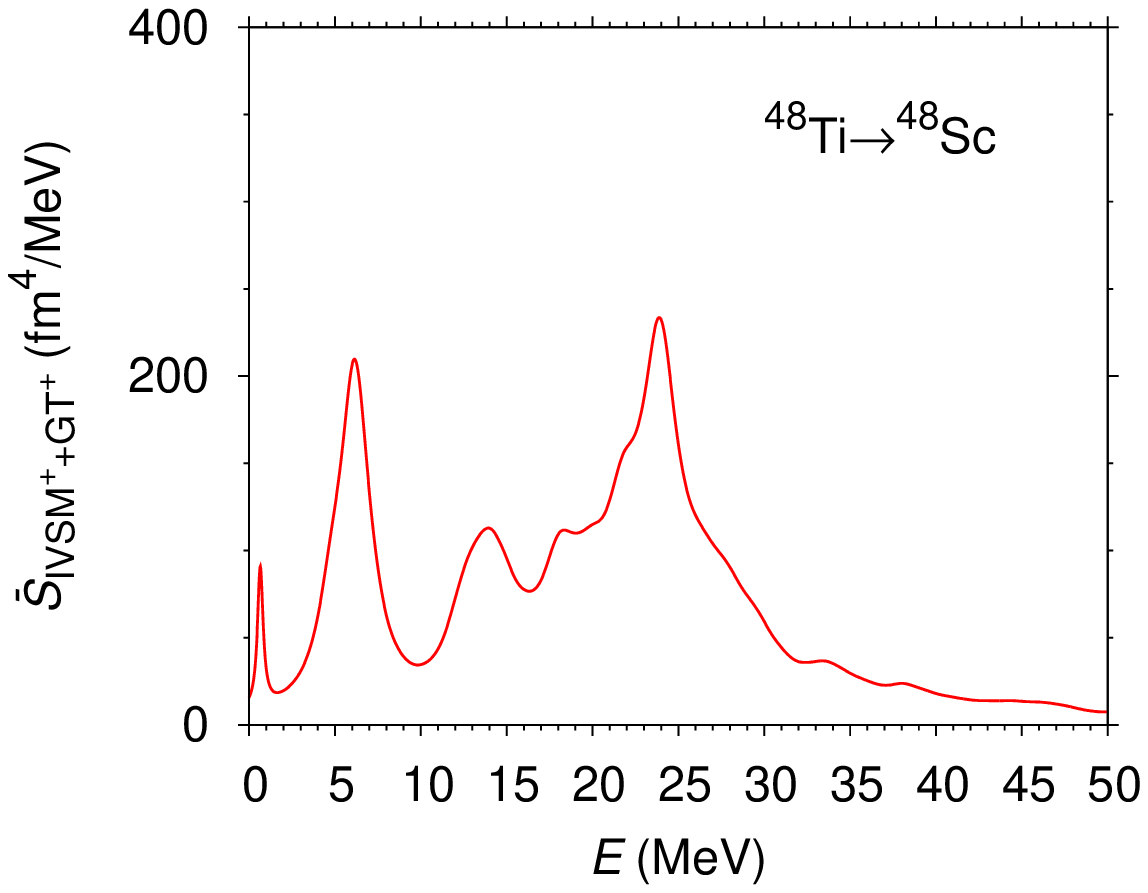}
\caption{ \label{fig:strfn_ivsm_48Ti} The same as Fig.~\ref{fig:strfn_ivsm_48Ca} but for $^{48}$Ti $\rightarrow$ $^{48}$Sc.}
\end{minipage}
\hspace{15pt}
\begin{minipage}[t]{.47\textwidth}
\includegraphics[width=1.0\columnwidth]{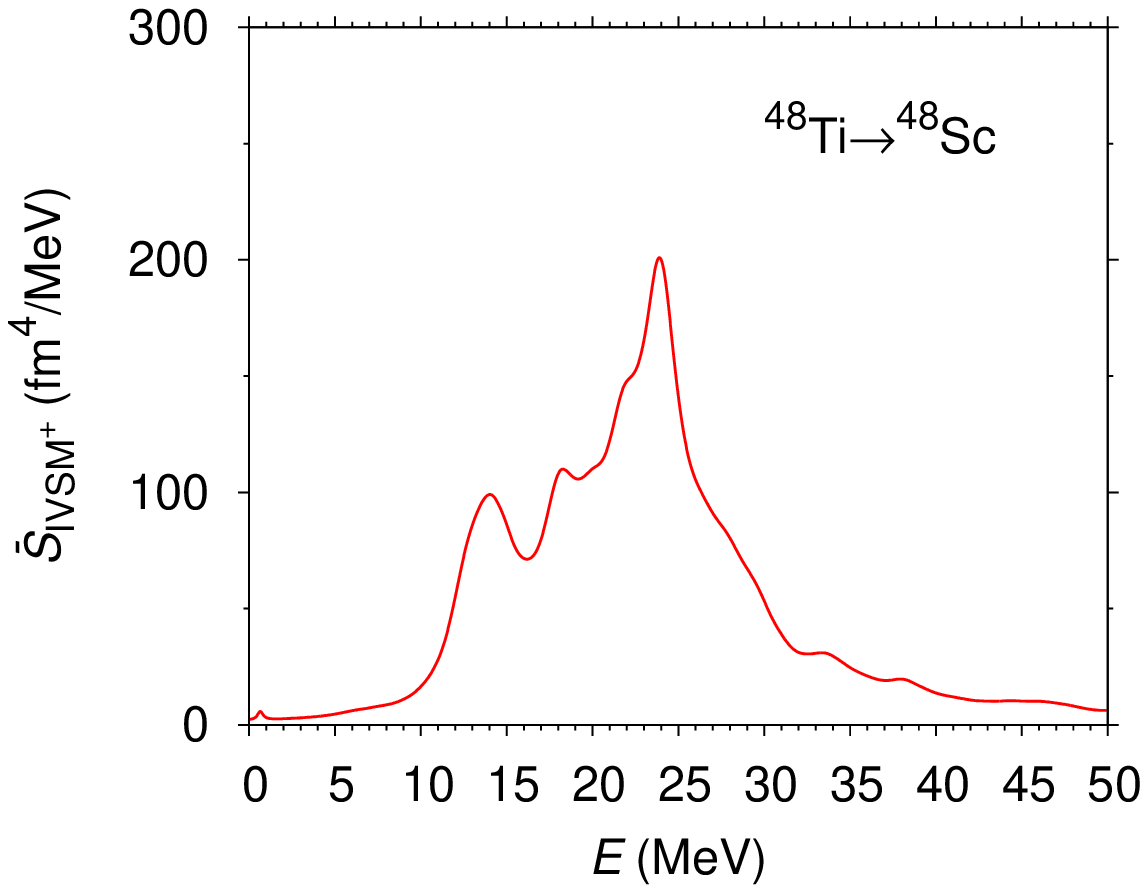}
\caption{ \label{fig:strfn_p2n_gt-sbtd_ivsm_48Ti} The same as Fig.~\ref{fig:strfn_gt-sbtd_ivsm_48Ca} but for $^{48}$Ti $\rightarrow$ $^{48}$Sc ($\langle r^2 \rangle$ of a proton 1$f_{7/2}$ is used).}
\end{minipage}
\end{figure*}

The experiments of $^{48}$Ca({\it p},{\it n})$^{48}$Sc and $^{48}$Ti({\it n},{\it p})$^{48}$Sc reactions have been made \cite{Yak09}, and the charge-exchange strength functions have been obtained. 
Figure \ref{fig:strfn_gt_48Ca} shows the measured and calculated GT strength functions for $^{48}$Ca $\rightarrow$ 
$^{48}$Sc. These results have two common structures; one is the low-energy peak ($E \simeq 2.5$ MeV in the experiment and 1.3 MeV in the calculation) and the giant resonance in $E=8$$-$13 MeV. There is no major structure above this energy region in the calculated result, but the tail of the experimental data is higher than the calculated one (see the inset). The width parameter of 0.2 and 1.0 MeV are used in the Lorenzian folding for the states below $\sim 8$ MeV and above this energy, respectively, for simulating the discrete and continuum states. The overall feature of the experimental data is reproduced. 
Figure \ref{fig:strfn_ivsm_48Ca} illustrates the strength function for $r^2\sigma\tau$. The same structure as that of the GT strength function is seen, in addition, a broad low distribution of the strength is seen in $E=15$$-$40 MeV. It can be speculated from the common feature of the two figures that the strength function for $r^2\sigma\tau$ also includes the GT strength, although the dimension is different. This speculation can be confirmed by Fig.~\ref{fig:strfn_gt-sbtd_ivsm_48Ca}, in which the transition operator $(r^2-\langle r^2\rangle_{n1f7/2})\sigma\tau$ is used \cite{Ham00}. The $\langle r^2 \rangle_{n1f7/2}$ is the mean square radius of one of the excess neutrons in 1$f_{7/2}$. 
The GT structure is almost removed, thus, the effectiveness of the separation method of Ref.~\cite{Ham00} is confirmed. On the basis of this physical interpretation, the strength function of Fig.~\ref{fig:strfn_ivsm_48Ca} and that of Fig.~\ref{fig:strfn_gt-sbtd_ivsm_48Ca} are denoted as 
$\bar{S}_{ \textrm{IVSM}^- + \textrm{GT}^- }(E)$ and 
$\bar{S}_{ \textrm{IVSM}^- }(E)$, respectively. 
Note that $S$ and $\bar{S}$ have the different dimensions. 
Figures \ref{fig:strfn_gt_48Ti}$-$\ref{fig:strfn_p2n_gt-sbtd_ivsm_48Ti} are those for $^{48}$Ti $\rightarrow$ $^{48}$Sc 
corresponding to Figs.~\ref{fig:strfn_gt_48Ca}$-$\ref{fig:strfn_gt-sbtd_ivsm_48Ca}. 
The calculated lowest-energy peak is lower than the corresponding experimental peak, and the calculated giant resonance ($E=4$$-$8 MeV) has more strength than the lowest-energy peak has. The corresponding experimental giant resonance around 6 MeV seems to be a shoulder. The operator of $r^2\sigma\tau$ yields the broad strength distribution  in $E \gtrsim 15$ MeV corresponding to that for $^{48}$Ca $\rightarrow$ $^{48}$Sc. 

There are two problems in $^{48}$Ca. 
It has been pointed out \cite{Yak09} that the shell-model calculation \cite{Hor07} does not have the high-energy broad distribution of the GT transition strength, although it has been argued in $^{90}$Zr by other groups (see the above brief review)  that the 2p-2h configurations create the corresponding strength distribution. The 2$\hbar\omega$-shell calculation of $^{48}$Ca \cite{Iwa15} does not show the broad strength distribution in the high-energy region, either. 
See also Refs.~\cite{Min16,Fuk17}, which use the second Tamm-Dancoff approximation, and Ref.~\cite{Cau90}. It is pointed out \cite{Dan97,Min16} that the tensor force has an effect to enhance the low and broad distribution of the GT transition strength in the high-energy region. Moreover, the reported $\mathcal{S}^\textrm{exp}_{\beta^-}$ is 15.3$\pm$2.2 ($E\leq 30$ MeV), which is 64$\pm$9 \% of the GT sum rule. That is, the quenching problem exists in $^{48}$Ca. 

\begin{figure*}[]
\begin{minipage}[t]{.47\textwidth}
\includegraphics[width=1.0\columnwidth]{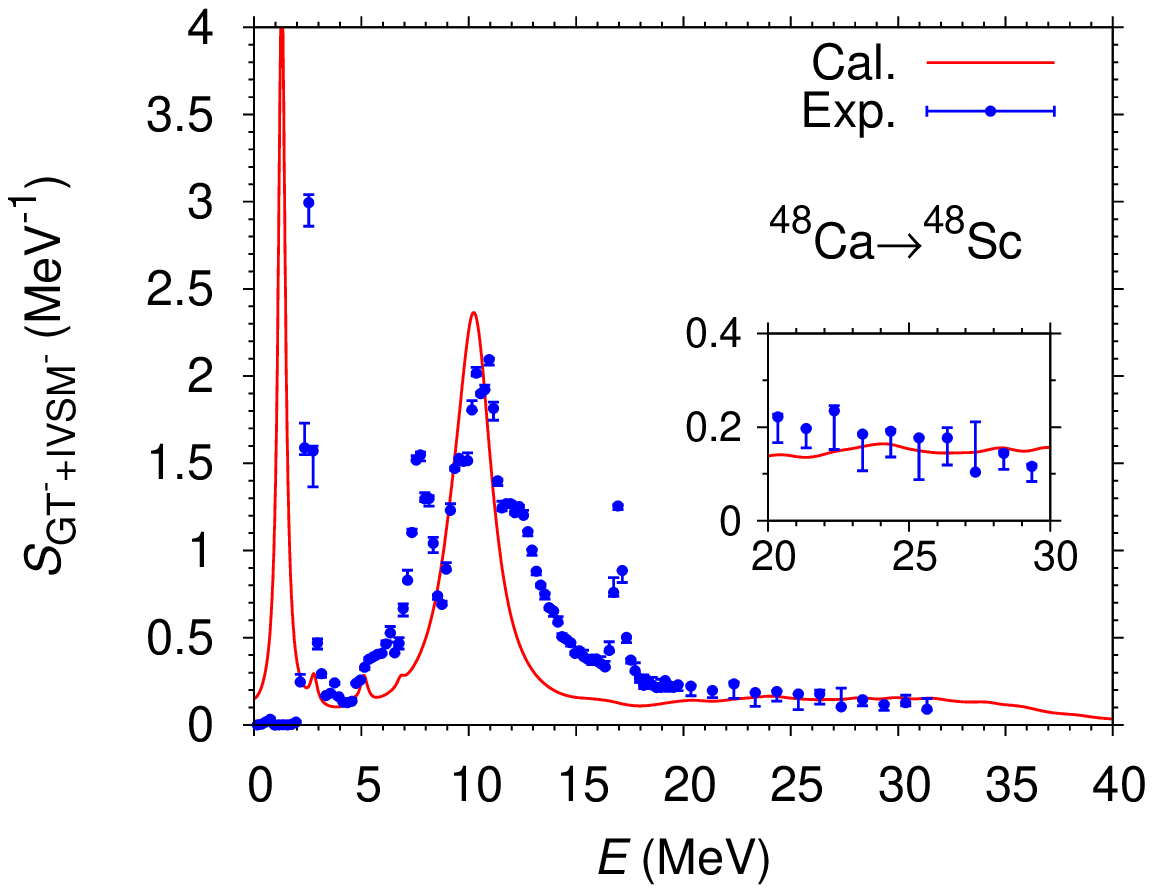}
\caption{ \label{fig:strfn_mix_48Ca} Strength functions measured (isolated points with error bars) and calculated using $O_\textrm{mix}$ [Eq.(\ref{eq:O_mix})] for $^{48}$Ca $\rightarrow$ $^{48}$Sc with $\alpha=-0.03$ fm$^{-2}$ (solid lines). The inset is a magnification.}
\end{minipage}
\hspace{15pt}
\begin{minipage}[t]{.47\textwidth}
\includegraphics[width=1.0\columnwidth]{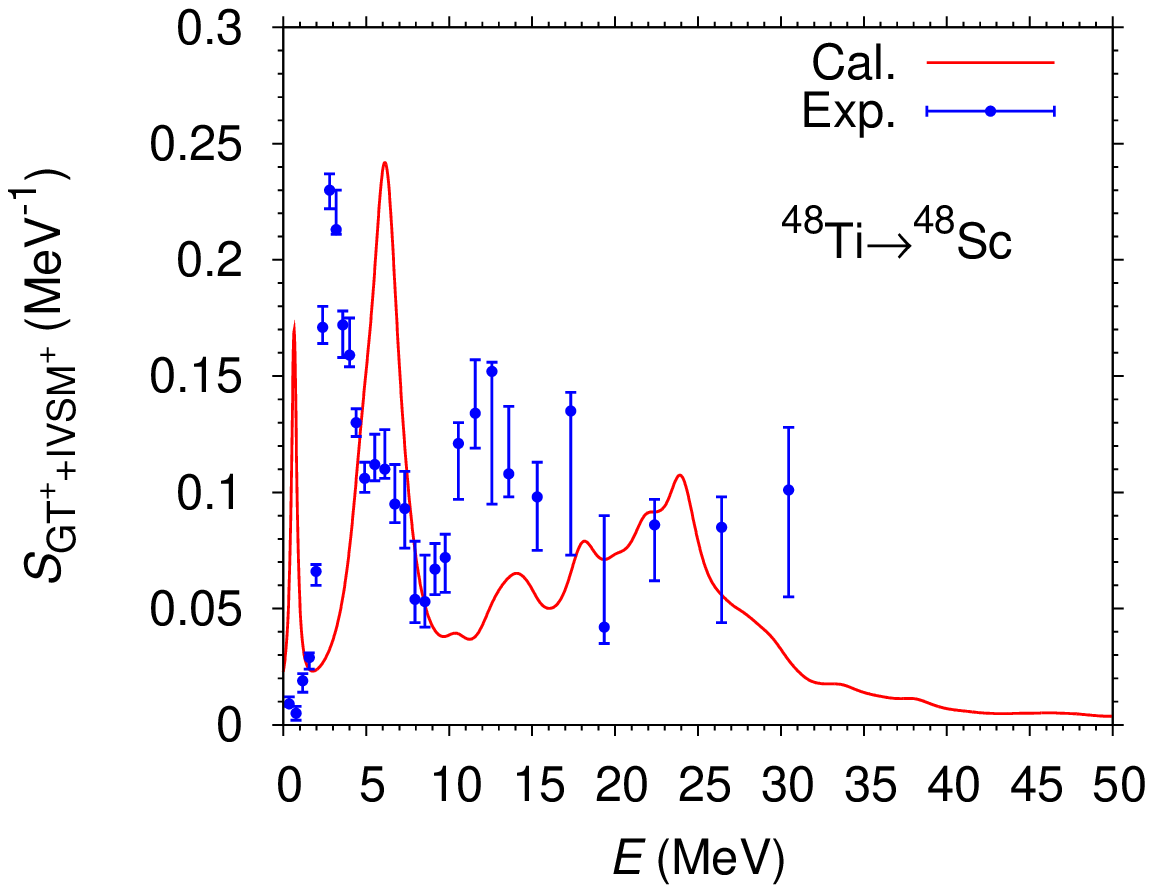}
\caption{ \label{fig:strfn_mix_p2n_48Ti} The same as Fig.~\ref{fig:strfn_mix_48Ca} but for $^{48}$Ti $\rightarrow$ $^{48}$Sc with $\alpha=- 0.0253$ fm$^{-2}$.}
\end{minipage}
\end{figure*}

I discuss these problems by introducing two hypotheses. 
The authors of Ref.~\cite{Yak09} state that their data contains the contribution of the isovector spin-monopole mode. 
From this information and the results of the calculations including the 2p-2h components mentioned above, I assume that the observed high-energy broad distribution of the strength is entirely due to the isovector spin-monopole mode for simplicity (the first hypothesis). 
The GT operator $\sigma\tau$ does not induce the transition strength of the isovector spin-monopole mode as shown by  Figs.~\ref{fig:strfn_gt_48Ca} and \ref{fig:strfn_gt_48Ti}. The tail of the calculated strength function in $E\gtrsim 20$ MeV of Fig.~\ref{fig:strfn_gt_48Ca} is the effect of the width parameter. A linear combination of $\sigma\tau$ and $r^2\sigma\tau$ is necessary for the transition operator causing both the GT and isovector spin-monopole components in the strength functions. 
Thus, I assume that the transition operator is 
\begin{align}
O_\textrm{mix} = \big( 1 + \alpha r^2 \big) \sigma\tau, \label{eq:O_mix}
\end{align}
where $\alpha$ is a constant having the dimension of the squared inverse length (the second hypothesis). An $r$-dependent extension of the transition operator is also studied in Ref.~\cite{Min16}. Because of the first hypothesis, this $\alpha$ can be determined phenomenologically using the QRPA wave functions. Since there is no GT strength in the calculations for $E \geq 15$ MeV ($^{48}$Ca $\rightarrow$ $^{48}$Sc) and $E \geq 10$ MeV ($^{48}$Ti $\rightarrow$ $^{48}$Sc), $\alpha$ can be determined so as to have the height of the experimental strength in those high-energy regions. 
Figures \ref{fig:strfn_mix_48Ca} and \ref{fig:strfn_mix_p2n_48Ti} show the results of this fitting. The negative $\alpha$'s (see the captions) are chosen because apparently those give the results close to the experimental data. The sum of the calculated strengths up to $E=30$ MeV is 
12.524 ($^{48}$Ca $\rightarrow$ $^{48}$Sc) and 2.243  ($^{48}$Ti $\rightarrow$ $^{48}$Sc), and the corresponding experimental values are  15.3$\pm$2.2 and 2.8$\pm$0.3, respectively. Both the calculated values are $\sim$ 20 \% smaller than the experimental values. The partial sums calculated with only the GT operator up to that energy (close to the saturated value) are  larger than the corresponding experimental values by $\sim$ 60 \% ($^{48}$Ca) and 14 \% ($^{48}$Ti). Thus, the problem of $^{48}$Ca is reduced significantly by the partial cancellation of the GT strength.  
For $^{48}$Ti, the result of calculation in Fig.~\ref{fig:strfn_mix_p2n_48Ti} is much closer to the experimental data than that in Fig.~\ref{fig:strfn_gt_48Ti}. 
It is an open problem why $^{48}$Ca and $^{90}$Zr are different in terms of the mechanism of the GT strength function. 
In Ref.~\cite{Mol15}, it is an implicit assumption that the isovector spin monopole component is not included in their charge-exchange data.  

\begin{figure*}[t]
\begin{minipage}[t]{.47\textwidth}
\includegraphics[width=\columnwidth]{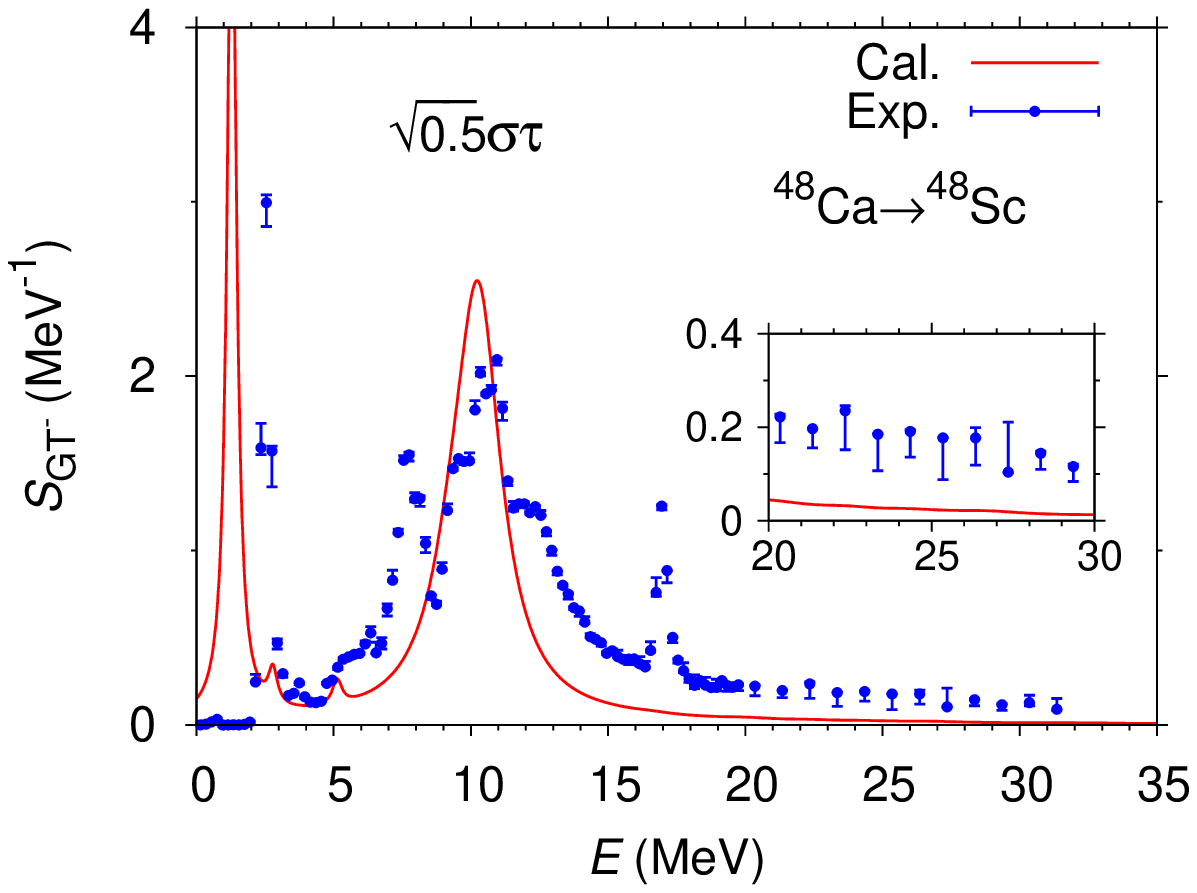}
\caption{ \label{fig:strfn_gt_rdcd0p50} The same as Fig.~\ref{fig:strfn_gt_48Ca} but for the quenched GT transition operator $\sqrt{0.5}\sigma\tau$.}
\end{minipage}
\hspace{15pt}
\begin{minipage}[t]{.47\textwidth}
\includegraphics[width=\columnwidth]{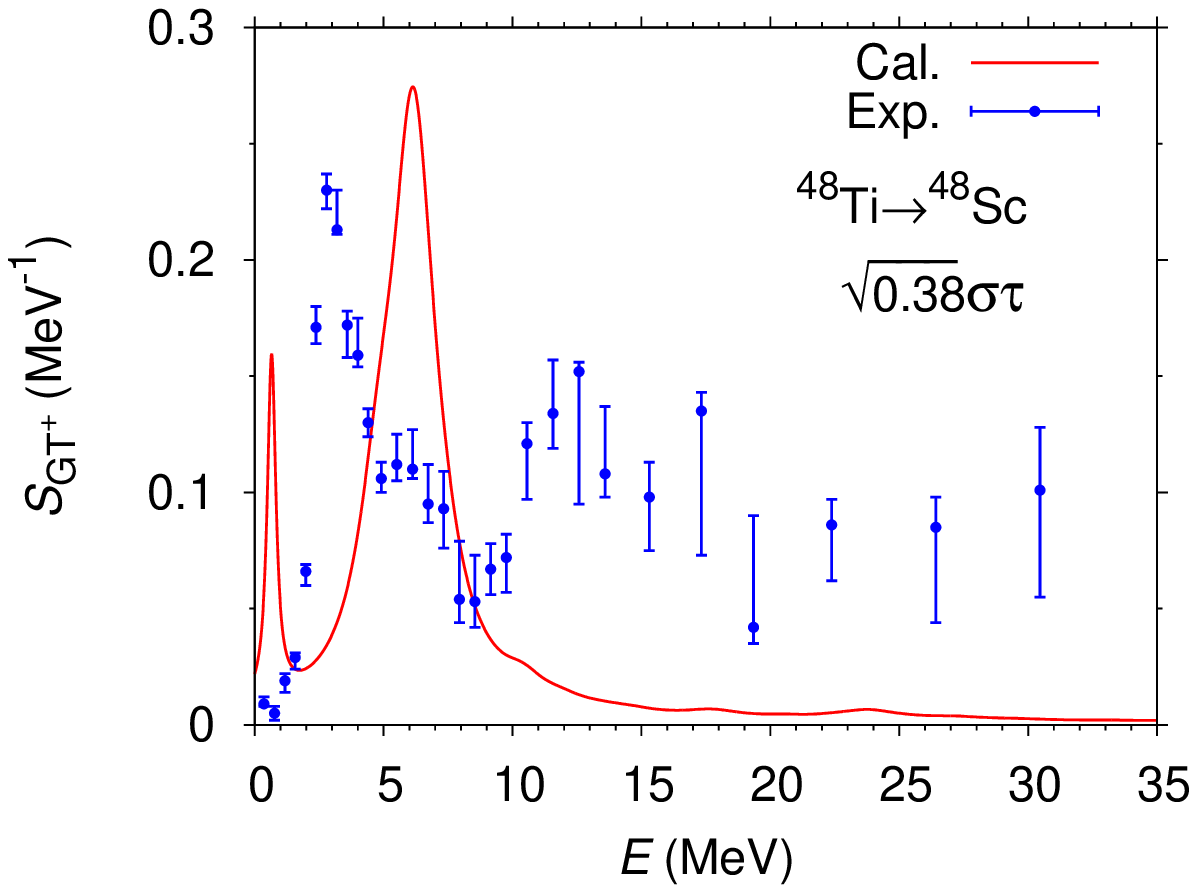}
\caption{ \label{fig:strfn_gt_p2n_rdcd0p38} The same as Fig.~\ref{fig:strfn_gt_48Ti} but for the quenched GT transition operator $\sqrt{0.38}\sigma\tau$.}
\end{minipage}
\end{figure*}

I also performed a reference calculation according to the usual phenomenology to multiply 
a quenching factor to the GT operator $\sigma\tau$, e.g, \cite{Alv04,Sen16,Cor17}. The results with $\sqrt{0.5}\sigma\tau$ 
($^{48}$Ca) and $\sqrt{0.38}\sigma\tau$ ($^{48}$Ti) are shown in Figs.~\ref{fig:strfn_gt_rdcd0p50} and \ref{fig:strfn_gt_p2n_rdcd0p38}, respectively. These quenching factors are chosen so as to reproduce approximately the sum of the experimental strengths up to 13 MeV ($^{48}$Ca) and 10 MeV ($^{48}$Ti). 
Using these quenched GT operators with $g_A^\textrm{bare}=1.27$ is equivalent to using effective $g_A=0.554$ in the calculation of $T^{(2\nu)}_{1/2}$; this $g_A$ is 15 \% larger than that of my calculation. 
The $T^{(2\nu)\textrm{th}}_{1/2}$ with this $g_A$ and $g^\textrm{pair}_{T=0}=-180.0$ MeV$\,$fm$^3$ is found to be 
3.76$\times$$10^{19}$ yr [Eqs.~(\ref{eq:nme2vGT_QRPA_I2})$-$(\ref{eq:nme2v_QRPA_I})] and  
4.12$\times$$10^{19}$ yr [Eqs.~(\ref{eq:nme2vGT_QRPA_F2})$-$(\ref{eq:nme2v_QRPA_F})]; 
these values are in the same order as that of the experimental data. 
Note the problems, however, 
that the isovector spin monopole strength in the high-energy region is ignored, and the GT sum rule is not satisfied, as I mentioned. 

\section{\label{sec:summary} Summary}
The $\beta\beta$ NMEs of $^{48}$Ca $\rightarrow$ $^{48}$Ti were calculated within the QRPA approach using the method developed recently \cite{Ter13,Ter15,Ter16}, and the consistency checks of my calculation have been made carefully. These check points are
\begin{enumerate}
\item the consistency of two sets of the intermediate-state energies obtained using the initial and final HFB states. This is a check for the $2\nu\beta\beta$ NME, and very close results were obtained. 
\item The consistency of the two decay paths of the $\beta\beta$ and two-like-particle transfer in the $0\nu\beta\beta$ GT NME under the closure approximation. 
This consistency was used for determining the strength of the $T=0$ pn pairing interaction.
\item The $2\nu\beta\beta$ Fermi NME is much smaller than the $2\nu\beta\beta$ GT NME, so that the isospin invariance of the $T=1$ pn pairing interaction is approximately satisfied. This was achieved by using the strength of the $(T,T_z)=(1,0)$ pairing interaction equal to the average of those of the $(T,T_z)=(1,1)$ and $(T,T_z)=(1,-1)$ pairing interactions.
\item The stability of the result with respect to the $T=0$ pn pairing interaction has been confirmed. 
\item The GT sum rule is satisfied well. 
\item The consistency with the experimental data of the GT strength function by the $(n,p)$ and $(p,n)$ reactions. I proposed a phenomenology explaining the quenched experimental data. 
\end{enumerate}
Check points 1 and 6 are new in the QRPA approach, and it should be possible to make these checks in the QRPA calculation generally. 
Check point 2 is an originality of my method. Check point 4 is related to this originality because referring to the lpQRPA, which does not use too strong interactions, has an effect to prevent too strong $T=0$ pn pairing interaction. Check points 3 and 5 are usual in the QRPA approach. The essence of the check point 5 is that enough large single-particle and two-particle spaces are used. 

My motivation to investigate $^{48}$Ca is to clarify whether this nucleus is particularly difficult to the QRPA approach. The possibility of this difficulty is in the pairing gaps of $^{48}$Ca, however, the uncertainty of these pairing gaps can be minimized by solving the HFB equation self-consistently. 

I used the enough large single-particle and two-particle spaces so that effective operator method for enhancing the short-range correlations as the Jastrow-like functions is not used.  
In Ref.~\cite{Ter15} I showed that the NME was almost saturated with respect to the increase in the two-particle space, and the same cutoff parameters were used in the present calculations. 
The $g_A$ was determined so as to reproduce the experimental half-life of the $2\nu\beta\beta$ decay. 
Because of that large single-particle space, I can use the same $g_A$ for the $0\nu\beta\beta$ and $2\nu\beta\beta$ decays. 
The only apparent disadvantage of the QRPA is that the low-lying $0^+$ excited state of $^{48}$Ca cannot be constructed. If there are excited states of  $^{48}$Sc obtained from that $0^+$ state by charge exchange, these states would not be included in the $\beta\beta$ NME of the QRPA approach. The qualitative reproduction of the experimental GT strength functions of $^{48}$Ca and $^{48}$Ti implies that this nuclear-structural problem does not affect the $J^\pi=1^+$ component of the GT NME. 
It is concluded that there is no clear problem in the QRPA approach to the $\beta\beta$ NMEs of $^{48}$Ca $\rightarrow$ $^{48}$Ti. 

The comparison of my result with those of other groups was made in terms of $R_{1/2}^{(0\nu)}$ and $M^{(0\nu)\prime}$.  
My result has the highest $R_{1/2}^{(0\nu)}$ and the lowest $M^{(0\nu)\prime}$ among the compared results and is close to one of other QRPA calculations. The QRPA approach usually has larger $M^{(0\nu)\prime}$ than the shell model has in many decay instances, however, exceptionally for $^{48}$Ca the QRPA has smaller values than the shell-model calculations. I obtained this result for $^{48}$Ca with a small $g_A$ in the present calculation. 

\begin{acknowledgments} 
I thank Dr.~Yako for giving me the data of the experiment. 
The numerical calculations of this paper were performed by 
the K computer at Advanced Institute for Computational Science, RIKEN through the program of High Performance Computing Infrastructure in 2016 (hp160052)  and 2017 (hp170288). Computer Coma at Center for Computational Sciences, University of Tsukuba was also used through
the Interdisciplinary Computational Science Program in 2016 (TKBNDFT) and 2017 (DBTHEORY). 
This study is supported by 
European Regional Development Fund-Project ``Engineering applications of microworld physics" (No.~CZ.02.1.01/0.0/0.0/16\_019/0000766). 
\end{acknowledgments}

\end{document}